\documentclass[a4paper,11pt]{article}
\pdfoutput=1
\usepackage{amssymb}   
\usepackage{amsmath}
 
\usepackage{graphicx}  
\usepackage{bm}        
\usepackage{dcolumn}   
\usepackage{jinstpub}
\usepackage{textcomp} 
\usepackage{color}

\setkeys{Gin}{clip=true,keepaspectratio=true} 
\begin{document}
   \title{High-Accuracy Absolute Magnetometry with Application to the Fermilab Muon $g-2$ Experiment}


\author[a,b]{D.~Flay,\footnote{Corresponding author, flay@jlab.org.}}
\author[a]{D.~Kawall,}
\author[c]{T.~Chupp,}
\author[d]{S.~Corrodi,}
\author[c]{M.~Farooq,}
\author[e,f,g]{M.~Fertl,}
\author[a]{J.~George,}
\author[d,c]{J.~Grange,}
\author[d]{R.~Hong,}
\author[e]{R.~Osofsky,}
\author[d]{S.~Ramachandran,}
\author[e]{E.~Swanson,}
\author[d]{and P.~Winter} 

\affiliation[a]{Department of Physics, University of Massachusetts, Amherst, MA 01003, U.S.A.}
\affiliation[b]{Experimental Nuclear Physics Division, Thomas Jefferson National Accelerator Facility, Newport News, VA 23606, U.S.A.}
\affiliation[c]{Department of Physics, University of Michigan, Ann Arbor, MI 48109, U.S.A.}
\affiliation[d]{High Energy Physics Division, Argonne National Laboratory, Lemont, IL 60439, U.S.A.}
\affiliation[e]{Center for Experimental Nuclear Physics and Astrophysics and Department of
Physics, University of Washington, Seattle, WA 98195, U.S.A.}
\affiliation[f]{Institute for Physics, Johannes Gutenberg University Mainz, Mainz, D-55128, Germany} 
\affiliation[g]{PRISMA+ Cluster of Excellence, Johannes Gutenberg University Mainz, Mainz, D-55128, Germany} 

   \abstract{
   We present details of a high-accuracy absolute scalar magnetometer 
   based on pulsed proton NMR. The $B$-field magnitude is determined 
   from the precession frequency of proton spins in a cylindrical 
   sample of water after accounting for field perturbations from 
   probe materials, sample shape, and other corrections. Features of the design, 
   testing procedures, and corrections necessary for qualification 
   as an absolute scalar magnetometer are described. The device was 
   tested at $B = 1.45$\,T but can be modified for a range exceeding 
   1--3\,T. The magnetometer was used to calibrate other NMR magnetometers 
   and measure absolute magnetic field magnitudes to an accuracy 
   of 19 parts per billion as part of a measurement of the muon 
   magnetic moment anomaly at Fermilab.
}


   \maketitle

\section{Introduction}


The accurate measurement of magnetic fields is a common requirement in atomic, 
nuclear, and particle physics experiments. However, most experiments have a unique 
set of requirements on field resolution, bandwidth, spatial resolution, 
field magnitude and measurement range, vector versus scalar measurement, 
and desired accuracy. To meet differing needs, many types of magnetometers have 
been developed and are in broad use, including Hall sensors, fluxgate magnetometers, 
rotating coil, optically-pumped alkali vapor magnetometers, Faraday rotation 
magnetometers, magnetoresistive devices, SQUIDS, and nuclear magnetic 
resonance (NMR) magnetometers.

For many experiments investigating the fundamental properties of particles, 
atoms, or molecules in magnetic fields, high accuracy absolute scalar magnetometers 
are required to determine the magnitude of the magnetic induction 
$B\left(\vec{r},t\right) = \left\vert \vec{B} \left( \vec{r},t \right) \right\vert$ in terms of 
a standard such as the Tesla. Ideally, all calibrated absolute scalar magnetometers 
should report the same value of $B$ when inserted in the same field, regardless of 
the kind of magnetometer.

The highest accuracy scalar magnetometers typically use the NMR signal of a 
nuclear spin precessing in an external $B$ field. The spin angular precession 
frequency $\omega_{N}$ of a bare nucleus $N$ is directly related to the external 
field magnitude $B$ through the Larmor relation $\omega_{N}=\gamma_{N} B$ where 
$\gamma_{N}$ is the gyromagnetic ratio of the bare nuclear spin in units of 
radians per second per Tesla. 

Since bare nuclei are difficult to work with, most practical magnetometers use 
nuclei shielded in an atom or molecule. One complication is that the field at 
the nucleus is reduced from the field to be measured by the diamagnetic shielding 
of the atomic electrons and other effects~\cite{Ramsey:1950}.  The reduction is 
directly proportional to the external field and of order 25\,parts per million (ppm) 
for protons in molecular hydrogen or water. For shielded nuclei, the Larmor relation 
is modified to $\omega'_{N}=\gamma'_{N}B$, where the shielded gyromagnetic ratios  
$\gamma'_{N}$ are typically less than the bare gyromagnetic ratios $\gamma_{N}$.
Despite this complication, the gyromagnetic ratios for shielded nuclei of practical 
interest for magnetometers, namely protons shielded in water $\gamma'_{p}$, and 
the $^{3}$He nucleus (helion) shielded in the $^{3}$He atom $\gamma'_{h}$, are known 
at the level of 11 and 12\,ppb respectively~\cite{Phillips:1977,Tiesinga:2018codata,Flowers:1993}. 
Thus in principle absolute field measurements can be made with these systems to 
11--12\,ppb absolute accuracy.  This should be compared with the limits of other 
common approaches such as Hall sensors, where the proportionality constant between 
the Hall voltage and the external field $B$ depends on the applied Hall current and 
sensor properties such as electron and hole densities and mobilities. These latter 
factors are not easily known or stable below the ppm level, setting a limit on 
the absolute accuracy.

The potential for high absolute accuracy of NMR magnetometers can only be realized 
with careful construction and analysis of signals.  For instance, the magnetization 
of materials used in the magnetometer construction will necessarily perturb the local field. 
When accuracies better than a part per million are sought, careful choice of materials 
and shapes is necessary to reduce this field perturbation, which must be measured. 
Trade-offs between absolute accuracy and resolution are also typically necessary. 
For instance, the RF coil used for detecting the precessing nuclear magnetic moments 
should be close to the NMR sample to maximize signal size and signal resolution. 
However, the magnetic field perturbation from the coil increases with proximity to 
the NMR sample, and fields from currents in the coil induced by the precessing nuclear 
magnetic moments, also perturbs the local field. Thus RF coil design is optimized 
differently for absolute magnetometry compared to magnetometers optimized for high precision. 
These and other design considerations, testing procedures, and corrections that should 
be considered for an absolute scalar magnetometer are discussed below.

\subsection{Overview of the Paper}

In Sec.~\ref{sec:design}, the design and performance of our calibration probe 
is presented.  The magnetic correction terms are identified in Sec.~\ref{sec:mag-cor}. 
In Sec.~\ref{sec:apps-hep}, we present applications of absolute magnetometry in particle 
physics.  The magnetic corrections for our magnetometer are quantified in Sec.~\ref{sec:mag-char}.  
An extensive cross-checking program against other absolute magnetometers is 
discussed in Sec.~\ref{sec:cross-check}.

\section{Design and Performance} \label{sec:design}  

Our design choices for the absolute magnetometer (hereafter referred to as the 
calibration probe) are guided by the desire to minimize the magnetic 
perturbation of the probe on the magnetic field.  A highly-symmetric construction 
using combinations of paramagnetic and diamagnetic materials minimize the magnetic 
perturbations. On the other hand, the NMR sample shape presents a non-negligible 
correction for $\omega_{p}'$.  Due to the difficulties associated with building 
an NMR sample with a spherical shape with aberrations far below the percent level, 
we use a cylindrical NMR sample shape since glass tubes with high symmetry are easily obtainable.   



\subsection{Operating Principles: Pulsed NMR} \label{sec:nmr} 

The probe is operated using pulsed NMR.  When placed in an external magnetic field,
proton magnetic moments build up a net polarization according to the Boltzmann distribution.  
A short-duration radio-frequency (RF) pulse at the resonant Larmor frequency is applied 
across the NMR sample using an excitation coil, generating an RF field perpendicular 
to the external magnetic field $\vec{B} = \vert B \vert \hat{y}$.  For instance, at the 
magnetic field $\vert B \vert = 1.45$\,T used for testing this probe, the resonant 
frequency is 61.79\,MHz.  At a specific duration, the RF pulse tips the magnetization 
of the proton ensemble perpendicular to $\vec{B}$, known as a $\pi/2$ pulse, see 
Fig.~\ref{fig:nmr}. The proton spins then precess in the horizontal $xz$ plane and 
induce excitation currents in the coil due to the changing magnetic flux.  The magnetization 
relaxes back to being aligned with $\vec{B}$, with time constant $T_{1}$.  
Simultaneously, spin-spin interactions and magnetic field gradients across the NMR 
sample cause these spins to dephase, damping out the oscillatory signal.  The time over 
which these effects occur is known as the $T_{2}^{*}$ time, written as: 

\begin{figure}[hbt]
   \centering
   \includegraphics[width=\linewidth]{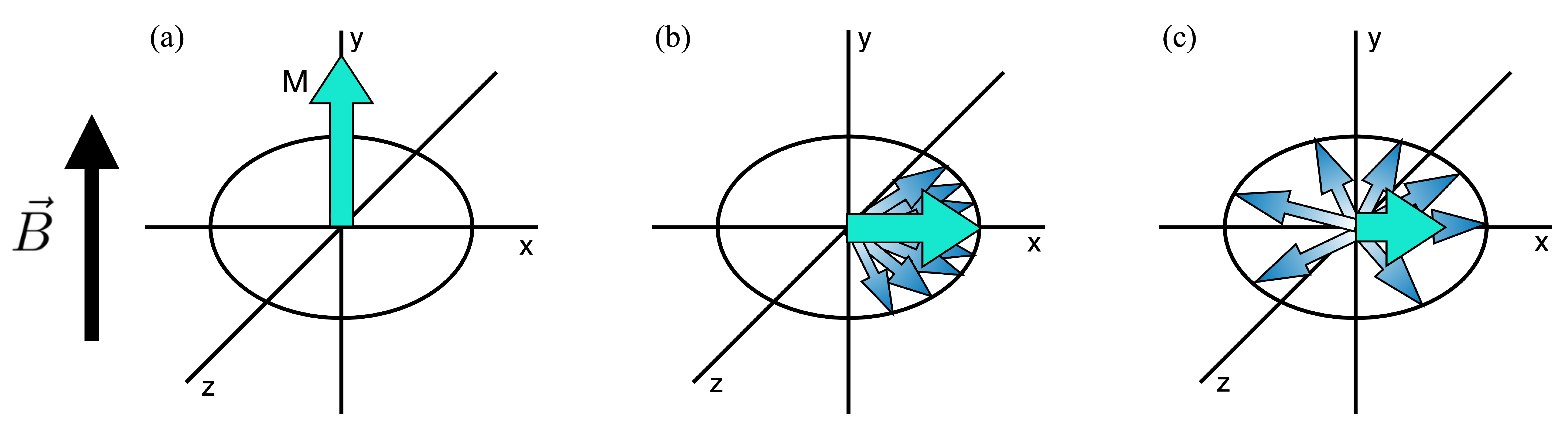}
   \caption{An illustration of the proton spin ensemble at various times 
            relative to the $\pi/2$ pulse.  The large arrow represents the 
            magnetization $M$ of the NMR sample and the smaller arrows represent individual proton spins.  
            Panel (a) is immediately before the $\pi/2$ pulse, panel (b) is shortly after, 
            and panel (c) is several seconds later.}
   \label{fig:nmr}
\end{figure}

\begin{equation}
   \frac{1}{T_{2}^{*}} = \frac{1}{T_{2}} + \frac{1}{T_{2}^{\dagger}}, 
\end{equation} 
 
\noindent where $T_{2}$ describes pure spin-spin interactions, and $1/T_{2}^{\dagger} \propto \gamma \Delta B$ 
denotes the decay time due to magnetic field gradients $\Delta B$ for an NMR sample with a 
gyromagnetic ratio $\gamma$.  The $T_{2}^{*}$ time is necessarily less than or equal to 
twice the $T_{1}$ time~\cite{Abragam:1961,Traficante:1991}.
This decaying oscillatory signal is known as the free-induction decay (FID) signal. 

To extract the frequencies of the probe's FIDs, we apply a zero-crossing counting
algorithm; the zero crossings and their neighboring data samples were fit to a line
to determine the $i^{\text{th}}$ crossing time $t_{0}^{i}$.  This in turn allows
the reconstruction of the phase of the FID signal as a function of time.  Fitting
the phase to a seventh-order odd-polynomial~\cite{Cowan_1996,Hong:2021hil} and taking the first derivative
evaluated at zero time gives the frequency of the FID.\footnote{Zero time corresponds to 
the start of the $\pi/2$ pulse when it is delivered to the probe.}  
In the analysis, the baseline of the signal is treated carefully; a time-varying 
baseline could potentially bias the extracted frequency.  We correct the baseline as 
follows: the algorithm first subtracts a constant baseline from 
the signal to center it on 0\,V.  The residual slope of the signal is then  
minimized by comparing neighboring pairs of zero crossings.  We compute an iterative 
correction by applying the Newton-Raphson method~\cite{Raphson:1697} to find the 
roots of the residual baseline as a function of the average time difference of all pairwise 
zero crossings.  The correction converges in about 4--5 iterations, and varied between 
0 and 1\,mV in magnitude.  The correction had a negligible impact on the extracted frequency 
of the FID.  The frequency analysis algorithm has been verified to be accurate to $<$1\,ppb 
when tested against a simulation of the probe's NMR signal.  
The simulation computes the voltage response of the NMR coil and accounts for its construction 
as well as the capacitance and inductance of the full resonant circuit (Sec.~\ref{sec:probe-elec}).  
The simulation also incorporates the probe's geometry and materials, and is performed 
in a variety of magnetic field gradients up to 20\,ppb/mm across the probe's 
active volume~\cite{Hong:2021hil}.  

\subsection{Calibration Probe Design} \label{sec:probe-design} 

\subsubsection{Mechanical Construction} \label{sec:probe-mech} 

The design of the probe is shown in Fig.~\ref{fig:pp-design}, and its physical assembly is 
given in Fig.~\ref{fig:pp-assembly}\footnote{See Appendix~\ref{sec:comp-listings} for a 
full list of the components used in the calibration probe.}.  
It features a near-zero magnetic susceptibility 
RF coil; it is a 0.97-mm outer-diameter (OD) copper tube filled with aluminum 
such that its effective magnetic susceptibility is 8\% that of copper.  The coil was 
wound to a length of 1\,cm with 5.5 turns and 2-mm pitch.  It is mounted on a 
$(15.065 \pm 0.008)$ mm OD, $(13.470 \pm 0.013)$ mm inner-diameter (ID) high-precision glass 
tube, with limits on concentricity of 38\,$\mu$m and camber of 13\,$\mu$m.  
This provides rigid, symmetric placement of the coil relative to the water sample, which 
is located at the center of the probe.  We utilize an ultra-pure ASTM Type-1 water 
sample, housed in a $(4.9635 \pm 0.0065)$ mm OD, $(4.2065 \pm 0.0065)$ mm ID glass tube that has 
limits on concentricity of 51\,$\mu$m and camber of 25\,$\mu$m.  
The tube is held at the center of the probe via insertion into a hole in the Macor\texttrademark{} 
support on the electronics end of the probe (left side of Fig.~\ref{fig:pp-design}). 
The opposite end of the water sample is held in place by a plastic adapter that slip-fits 
inside the 8-mm opening of the Macor\texttrademark{} support at the far end of the probe 
(right side of Fig.~\ref{fig:pp-design}). The Macor\texttrademark{} parts have machining 
tolerances of 0.05\,mm. The RF coil leads travel down opposite sides of 
the glass support tube, pass through 2-mm diameter holes in the Macor\texttrademark{} support, 
and are soldered to an electronics board.   The outer shell of the probe is 25.4-mm OD, 
1.0-mm wall aluminum alloy 2024.  This shell is connected to the ground of the 
probe's circuit via a copper wire, see Fig.~\ref{fig:pp-assembly}.  

\begin{figure}[hbt]
   \centering
   \includegraphics[width=\linewidth]{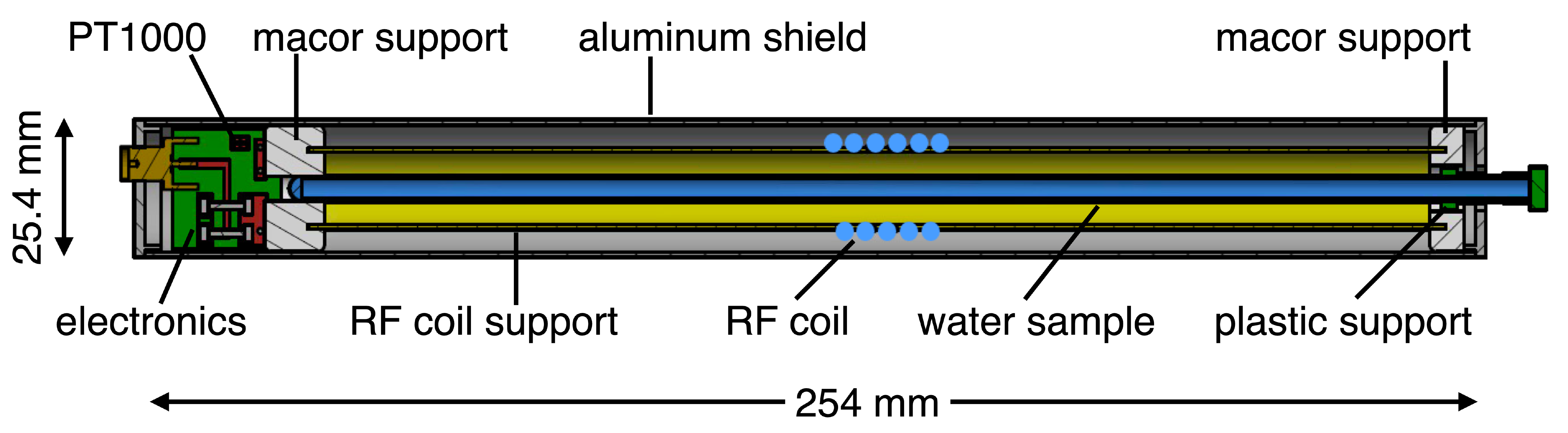}
   \caption{Design drawing for the calibration probe.  Tunable capacitors and a temperature 
            sensor are housed in the section at the far left end of the device.  The straight 
            leads of the RF coil are not drawn.  Figure reproduced from Ref.~\cite{Albahri:2021kmg}.}
   \label{fig:pp-design}
\end{figure}

\begin{figure}[hbt]
   \centering
   \includegraphics[width=\linewidth]{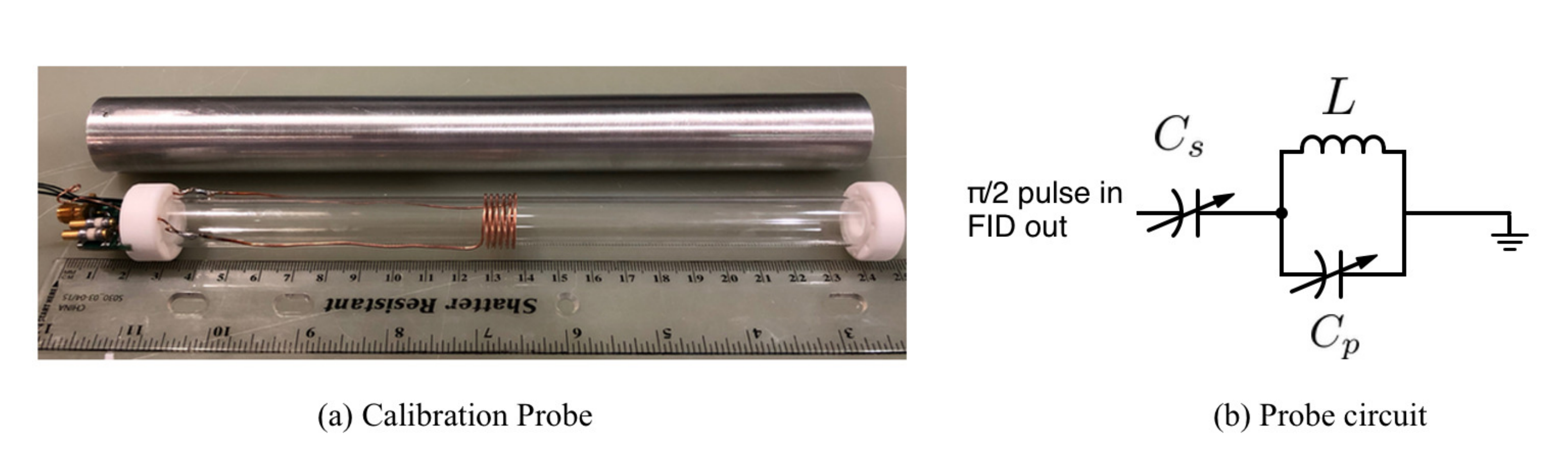}
   \caption{(a) The internal assembly of the probe with its aluminum shield.  The RF coil and its leads 
            are visible, along with the electronics board on the left.  
            (b) A schematic depiction of the probe circuit. The series and parallel 
            tunable capacitances $C_{s}$ and $C_{p}$, and coil inductance $L$ are indicated.  
            The probe shell is also connected to ground (not drawn). 
            }
   \label{fig:pp-assembly}
\end{figure}

\subsubsection{Electrical Circuitry} \label{sec:probe-elec}

The RF coil, which has an inductance of 0.5\,$\mu$H, is connected in parallel 
to a variable capacitor with a range of 1--12\,pF.  
This LC combination is connected in series to another variable capacitor of the same 
range.  This series capacitor is then connected in series to an SMA connector 
with the cable that attaches to the electronics unit that delivers the $\pi/2$ pulse 
and receives the FID signal (described in the following section).  
The physical implementation of the circuitry is shown in Fig.~\ref{fig:pp-assembly}, 
where all components are mounted on a 22-mm wide by 18-mm long electronics PC board.  
A PT1000 temperature sensor is mounted on the reverse side of the board for 
temperature monitoring, and was determined to be in thermal equilibrium with the water sample.  
The PT1000 is read out over four wires by a digital multimeter (DMM). 
The DMM was calibrated against a precision 1\,k$\Omega$ resistor, and was operated with 
a readout range of 10\,k$\Omega$ to minimize self-heating effects. 
When comparing various PT1000 sensors against one another, we found the sensor 
stability to be better than $\pm 0.5^{\circ}$C.  

\subsection{NMR Data Acquisition and Readout} \label{sec:daq}
 
In order to read out the calibration probe NMR signals, we designed an 
NMR data acquisition system.  The schematic design for the system is shown 
in Fig.~\ref{fig:spu-schematic}\footnote{See Appendix~\ref{sec:comp-listings} for a 
full list of the components used in the data acquisition system.}.  
The $\pi/2$ pulse is provided by a frequency synthesizer (F1) and a 250-W 
amplifier (A1).  This amplified signal (up to 10\,W, with durations between 30--50\,$\mu$s) 
is delivered to the NMR probe through a single-pole, double-throw RF switch (SW).  
The FID signal (typically 50--75\,$\mu$V in amplitude) is acquired 
by toggling the RF switch to a pre-amplifier (A2).  After amplification, this signal 
is band-pass filtered (BP, $(61.79 \pm 2.5)$\,MHz), and mixed down to $\approx$10\,kHz (MX).  
The local oscillator signal for the mixer is provided by another frequency synthesizer (F2).  
Both F1 and F2 use the same clock input, a GPS-disciplined 10\,MHz reference clock.  
The mixed-down FID is sent through additional filtering (LP1 and LP2, with cutoff frequencies of 
130\,kHz and 35\,kHz, respectively) and amplification (OA with a gain of 
$\approx$28\,dB).  The signal amplitude is $\approx$1\,V, which is digitized at 
10\,MHz (DG).   

\begin{figure}[hbt]
   \centering
   \includegraphics[width=\linewidth]{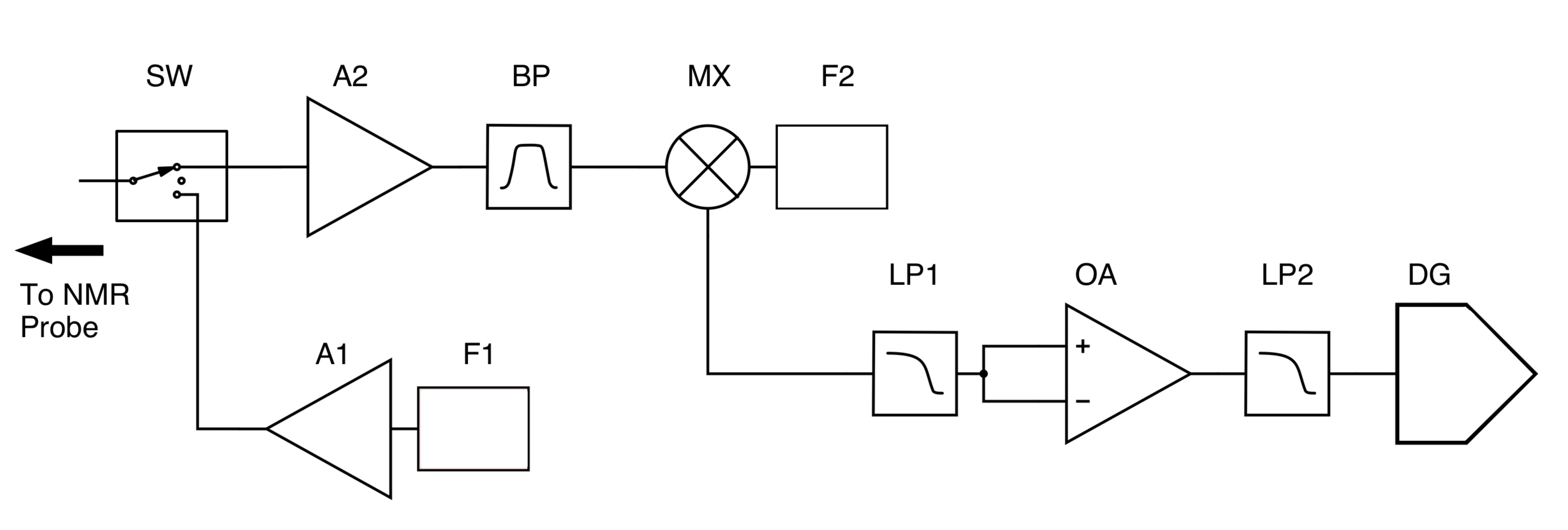}
   \caption{Signal processing schematic diagram.  Item SW is an RF switch.  
            The amplifiers are indicated by A1 and A2.  
            Frequency synthesizer units are indicated by F1 and F2.  
            The band-pass filter is indicated by BP and low-pass filters 
            LP1 and LP2. The frequency mixer is indicated by MX. 
            The operational amplifier is denoted by OA, and the digitizer is indicated by DG.
            For more details, see text.  
           }
   \label{fig:spu-schematic}
\end{figure}

In order to operate the system, we utilize a number of transistor-transistor logic pulses.  These signals 
are obtained via a field-programmable gate array (FPGA) unit with timing precision at the nanosecond scale, 
exceeding our requirements of microsecond-scale timing.  The pulse characteristics 
are managed by data acquisition software written in C++ and running on Linux.

The power required for the operation of the SW, A2, and OA components is provided by a separate 
power supply unit, which houses a 3.3\,V, $\pm$5\,V, and 12\,V linear power supplies.  
The digitizer DG and FPGA units are powered and operated through a VME interface. 

\subsection{Performance} \label{sec:performance}

\begin{figure}[hbt]
   \centering
   \includegraphics[width=\textwidth]{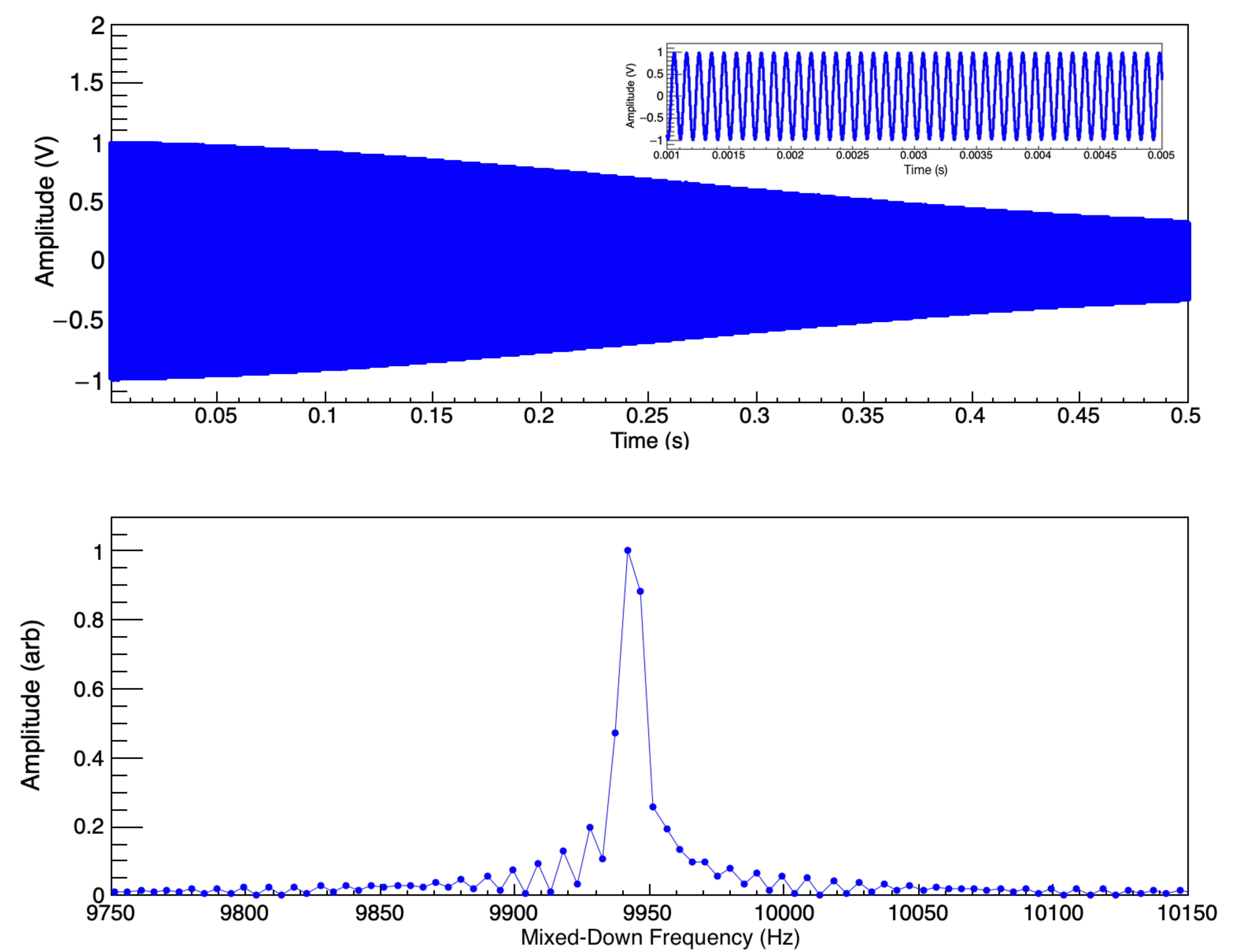}
   \caption{A typical FID recorded by the calibration probe when placed in the test  
            magnet at ANL.  Top panel: The NMR signal in the time domain; the inset 
            shows a zoomed-in portion of the signal.  Bottom panel: Fourier transform 
            of the signal, with its spectrum normalized to its peak value.  The frequency 
            scale shown is that of the mixed-down frequency.   
           }
   \label{fig:pp-signal}
\end{figure}

The probe has peak amplitudes of $\approx$1\,V with a noise
baseline of $\approx$5.5\,$\mu$V/$\sqrt{\text{Hz}}$ over a bandwidth of 35\,kHz  
for signal-to-noise ratios of $\approx$1000.  A typical FID signal and its Fourier transform 
are shown in Fig.~\ref{fig:pp-signal}.  In our test magnet facility at Argonne National Laboratory (ANL) 
and in-situ at Fermilab, we have achieved signal lengths of at least 400\,ms, 
where the amplitude of the signal drops to $1/e$ of its initial maximum value. 

\begin{figure}[hbt]
   \centering
   \includegraphics[width=\textwidth]{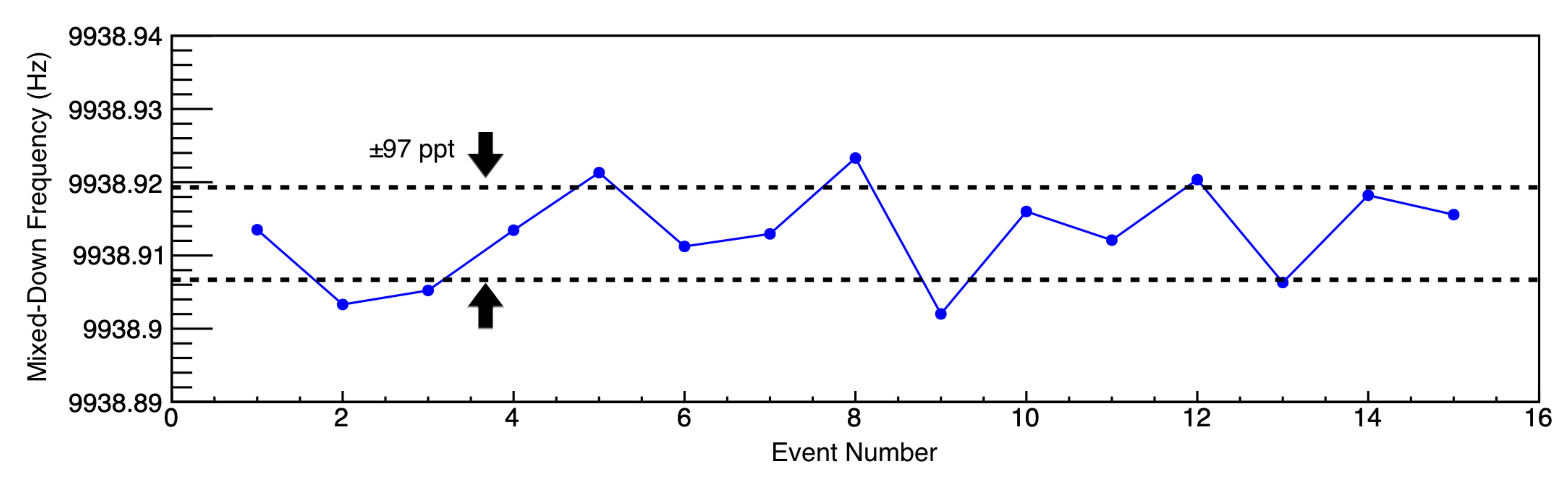}
   \caption{A typical data set showing the sensitivity of the calibration probe at ANL in 
            the test magnet.  The data acquisition system is triggered once every 5\,seconds.  
            The one-sigma band is indicated by the dashed lines, representing a standard 
            deviation of 97\,ppt.}
   \label{fig:pp-shot-res}
\end{figure}

The testing of the probe was performed at ANL at a test solenoid facility, which uses a 
large-bore magnetic resonance imaging (MRI) superconducting magnet in persistent mode.  
In this magnet, we have achieved a frequency resolution of $\lesssim$100\,parts-per-trillion (ppt) 
per single FID, due in part to our abilities to shim the MRI solenoid magnet to 
extremely high uniformity with gradients at the $\approx$1\,ppb/mm level, and to measure 
magnetic field drifts of $\approx$9\,ppb/hr.  A typical data set is shown in 
Fig.~\ref{fig:pp-shot-res}. In the Fermilab Muon $g-2$ magnet, the achievable magnetic 
field gradients are 10--20\,ppb/mm and the field stability is not as good, which reduces the 
single-FID frequency resolution to 10\,ppb on average.

\section{Magnetic Perturbations and Their Corrections} \label{sec:mag-cor}  

An NMR probe is sensitive not only to the field it is immersed in, but also
to the various magnetic aspects of the probe itself.  This includes the NMR
sample molecular structure and physical shape, and the probe material composition,
geometry, and radiation damping.  The magnetic field written in terms of the NMR
frequency of the protons in the sample is:\footnote{The magnetic perturbation
corrections are, in principle, multiplicative; we use the linear approximation
$\left(1 + \delta_{1}\right)\left(1 + \delta_{2}\right) \approx 1 + \delta_{1} + \delta_{2}$
since the $\delta$ terms are $\mathcal{O}$(ppm) or less, and hence higher-order terms are negligible.}

\begin{equation} \label{eqn:omega-p-shield}
   \omega_{p}'(T) = \omega_{p}^{\text{cp}}(T) \times \left[ 1 + \delta^{b}\left( \text{H}_{2}\text{O}, T \right) + \delta^{t} \right],
\end{equation}

\noindent where $\omega_{p}^{\text{cp}}(T)$ denotes the measured frequency of the
calibration probe at a temperature $T$, $\delta^{b}\left( \text{H}_{2}\text{O}, T \right)$
the bulk magnetic susceptibility of water, and $\delta^{t}$ the corrections to the field
due to the probe materials and other effects:

\begin{equation*}
   \delta^{t} = \delta^{s} + \delta^{p} + \delta^{\text{RD}} + \delta^{d}. 
\end{equation*}

\noindent The quantity $\delta^{s}$ characterizes corrections due to the probe material
and geometry, $\delta^{p}$ represents the correction due to paramagnetic impurities
in the water sample and asymmetries in the water sample holder, $\delta^{\text{RD}}$ and
$\delta^{d}$ correspond to dynamic effects relating to radiation damping and proton
dipolar fields, respectively. Each of these terms will be discussed in Sec.~\ref{sec:mag-char}.

The bulk magnetic susceptibility $\delta^{b}$ depends on the geometry and volume magnetic 
susceptibility of the NMR sample, given as:

\begin{equation} \label{eqn:bulk-mag-suscept} 
   \delta^{b}\left(\text{H}_{2}\text{O}, T\right) = \left( \varepsilon - \frac{1}{3} \right)\chi\left(\text{H}_{2}\text{O},T\right), 
\end{equation}

\noindent where $\varepsilon$ denotes the shape factor of the NMR sample, given as 
$1/3$ for a perfect sphere and $1/2$ for a perfect and infinite cylinder with its 
long axis perpendicular to $\vec{B}$~\cite{Osborn:1945}.  The volume magnetic susceptibility 
$\chi$ for water has been parameterized based on a compilation of measurements~\cite{Philo:1980}:

\begin{equation} \label{eqn:water-mag-suscept} 
   \chi\left( \text{H}_{2}\text{O},T\right) = \chi\left( \text{H}_{2}\text{O},20^{\circ}\text{C}\right) 
                                            \times  \left[ 1 + a_{1} \left( T - 20^{\circ}\text{C} \right) 
                                            + a_{2} \left(T - 20^{\circ}\text{C} \right)^{2} 
                                            + a_{3} \left(T - 20^{\circ}\text{C} \right)^{3} \right],  
\end{equation}

\noindent where $\chi\left(\text{H}_{2}\text{O},20^{\circ}\text{C}\right) = -9032 \times 10^{-9}$~\cite{Schenck:1996}.
We assign an uncertainty of $30 \times 10^{-9}$ by comparing with a measurement taken at
an unknown temperature, $\chi\left(\text{H}_{2}\text{O}\right) = -9060(3) \times 10^{-9}$~\cite{Blott:1993}.
The terms $a_{i}$ are
$a_{1} = 1.39 \times 10^{-4}/^{\circ}\text{C}$, $a_{2} = -1.27 \times 10^{-7}/\left(^{\circ}\text{C}\right)^{2}$,
and $a_{3} = 8.09 \times 10^{-10}/\left(^{\circ}\text{C}\right)^{3}$~\cite{Philo:1980}.

\subsection{The Free-Proton Larmor-Precession Frequency} \label{sec:free-prot}  

To extract the free proton precession frequency, there is a temperature-dependent
diamagnetic shielding correction $\sigma$:

\begin{equation} \label{eqn:omega-p-free} 
   \omega_{p}^{\text{free}}(T) = \omega_{p}'(T) \times \left[ 1 + \sigma\left( \text{H}_{2}\text{O},T \right) \right],  
\end{equation}

\noindent where $\sigma$ takes the form:

\begin{equation} \label{eqn:water-diamag-shield}
   \sigma \left( \text{H}_{2}\text{O},T \right) = \sigma \left( \text{H}_{2}\text{O}, 25^{\circ}\text{C} \right) +  
                                                  \frac{d\sigma \left(\text{H}_{2}\text{O}\right)}{dT} \left( 25^{\circ}\text{C} - T \right). 
\end{equation}

\noindent The quantity $\sigma \left( \text{H}_{2}\text{O},25^{\circ}\text{C} \right) = 25\text{ }691(11) \times 10^{-9}$~\cite{Mohr:2012tt}
and $d\sigma\left(\text{H}_{2}\text{O}\right)/dT = -10.36(30) \times 10^{-9}/^{\circ}\text{C}$~\cite{Petley:1984,Phillips:1977,Neronov:2014}.

\section{Applications of Absolute Magnetometry in Particle Physics} \label{sec:apps-hep}
 
Proton NMR has been used in muonium hyperfine experiments at Los Alamos~\cite{Liu:1999iz} 
and the Brookhaven National Lab (BNL) E821 Muon $g-2$ Experiment~\cite{Bennett:2004pv,Bennett:2006fi}.  
This absolute magnetometer used the pulsed-NMR technique (Sec.~\ref{sec:nmr}), and featured 
a spherical water sample encased in a long cylindrical aluminum shield.  This device 
achieved an accuracy of 34\,ppb~\cite{Fei:1997sd}. The leading limiting factors in 
its performance were the magnetic perturbation of its materials and the asphericity 
of the spherical glass water sample holder.      

A similar magnetometer design will be used in the upcoming experiment MuSEUM 
at J-PARC~\cite{Torii:2015sra}, but features a cylindrical water sample~\cite{Yamaguchi:2019iwm}. 
This absolute magnetometer uses continuous-wave (CW) NMR and has been evaluted 
to be accurate to 18\,ppb, where the uncertainty is dominated by the magnetic 
perturbations of the materials used in the magnetometer.   

Absolute magnetometry is not limited to using water-based NMR samples; a $^{3}$He-based 
absolute magnetometer that uses pulsed NMR has been constructed recently~\cite{Farooq:2019phd,Farooq:2020swf}.  
Studies presented in Ref.~\cite{Farooq:2020swf} show the $^{3}$He magnetometer 
agrees with the BNL E821 water-based magnetometer to within 32\,ppb when placed 
in the same magnetic field.  The uncertainty is dominated by corrections due 
to the materials of the $^{3}$He magnetometer.

\subsection{Overview of Magnetic Field Measurements for Muon $g-2$ Experiments}

NMR has been used to quantify the magnetic field in a number of muon $g-2$ experiments, 
including BNL E821 in the early 2000s~\cite{Bennett:2004pv,Bennett:2006fi}, 
the ongoing experiment at Fermi National Accelerator Laboratory (Fermilab) E989~\cite{Abi:2021gix}, 
and will be used for the upcoming experiment under construction at J-PARC~\cite{Abe:2019thb}. 

The magnetic moment of the positive muon is written: 

\begin{equation*}
   \vec{\mu}_{\mu} = g_{\mu} \left( \frac{e}{2m_{\mu}} \right) \vec{s}, 
\end{equation*}

\noindent where $e$ denotes the electric charge of the muon, $m_{\mu}$ its mass, 
$\vec{s}$ its spin, and $g_{\mu} = 2\left( 1 + a_{\mu}\right)$.  The quantity $a_{\mu}$ 
is the muon magnetic anomaly, representing radiative corrections due to interactions 
of the muon with virtual fields in the quantum-mechanical vacuum.  There is a 4.2$\sigma$ 
discrepancy between the theoretical prediction~\cite{Aoyama:20201} and the experimental 
measurements~\cite{Abi:2021gix,Bennett:2006fi}, hinting at new physics beyond the 
Standard Model.  

The quantity $a_{\mu}$ is determined experimentally as a ratio of two angular frequencies. 
The intensity variation of high-energy positrons from muon decays encodes the difference 
between the muon spin precession and cyclotron frequencies in the magnetic field of 
a storage ring denoted by $\omega_{a}$.  The storage ring magnetic field magnitude $B$ is measured using 
proton NMR and calibrated in terms of the spin precession frequency of protons shielded 
in a spherical water sample $\omega_{p}'$ at a reference temperature $T_{r} = 34.7^{\circ}$C.  
The quantity $a_{\mu}$ is extracted by combining these measurements with the 
quantities $\mu_{p}'(T_{r})/\mu_{e}(H)$, $\mu_{e}(H)/\mu_{e}$, $m_{\mu}/m_{e}$~\cite{Phillips:1977,Liu:1999mu,Tiesinga:2018codata} 
and $g_{e}$~\cite{Gabrielse:2011ge}: 

\begin{equation*}
   a_{\mu} = \frac{\omega_{a}}{\tilde{\omega}_{p}'(T_{r})} \frac{\mu_{p}'(T_{r})}{\mu_{e}(H)} \frac{\mu_{e}(H)}{\mu_{e}} \frac{m_{\mu}}{m_{e}} \frac{g_{e}}{2},  
\end{equation*}  

\noindent where the tilde indicates that the magnetic field is weighted by the muon 
beam intensity distribution across the cicular cross section of the toroidal muon 
storage volume and averaged over the storage ring azimuthal angle. 
 
The previous muon $g-2$ experiment (BNL E821) and the current experiment (Fermilab E989) 
utilize the same 14.2-m diameter superconducting storage ring magnet that produces a 1.45\,T magnetic 
dipole field across its 18-cm gap~\cite{Danby:2001eh}.  The magnetic field around the ring 
is mapped using a motorized cart dubbed the ``trolley''~\cite{Corrodi_2020}, which houses 
17 NMR probes that contain petroleum jelly NMR samples.  A single magnetic field map takes 
roughly 70 minutes.  The muon beam is stopped every 2--3 days for this measurement.  
While the muon beam is circulating in the superconducting magnetic storage ring, the field 
is continuously monitored by ``fixed'' NMR probes installed in grooves machined into the 
walls of the main vacuum chamber above and below the muon storage region. These fixed probe 
measurements are tied to those of the trolley during the trolley maps, and are subsequently 
used to track magnetic field changes over time in the beam storage region.  

To establish the absolute scale of the magnetic field seen by the trolley, a well-understood 
standard calibration probe is required. The magnetic characteristics of the calibration 
probe are measured accurately so they can be accounted for in the frequency measurements 
to extract the shielded precession frequency of protons in a water sample, 
$\omega_{p}'$.  A dedicated comparison of measurements between the standard calibration probe 
and the trolley in-situ at Fermilab is performed to transform trolley measurements into 
calibrated magnetic field maps.  This program produces a correction for each 
trolley probe to account for the magnetic perturbation of the trolley on the field 
measurements~\cite{Albahri:2021kmg}.  For Fermilab E989, the procedure entails the 
in-vacuum comparison of the trolley magnetic field measurements to those of the calibration 
probe at a specific location in azimuth in the storage ring.  As such, the calibration probe 
must be vacuum compatible, necessitating the use of low-outgassing materials in the 
probe construction (see Sec.~\ref{sec:design}).  

The level of precision achieved for the magnetic field calibration in BNL E821 was 
90\,ppb~\cite{Bennett:2006fi}, with a separate 50\,ppb attributed to the calibration probe~\cite{Fei:1997sd}.  
The total uncertainty budget for the magnetic field measurements in E989 is 70\,ppb, 
with 35\,ppb allotted to the calibration probe~\cite{Grange:2015fou}.  We have designed, 
built, and deployed a calibration probe in Fermilab E989 with an accuracy of 15\,ppb. 
 
\section{Magnetic Characteristics of the High-Accuracy Calibration Probe} \label{sec:mag-char}  

The magnetic characteristics of the calibration probe affect the field experienced 
by the protons in the water sample.  The correction $\delta^{t}$ associated with 
this must be quantified to high precision in order to accurately extract the shielded proton  
Larmor frequency $\omega_{p}'$.  The magnetic effects are categorized 
into intrinstic and configuration-specific terms, discussed in detail in the following 
subsections.    

\subsection{Intrinsic Effects}  \label{sec:intrinsic-effects} 

Intrinsic effects relate to the perturbation of the external magnetic field due 
to the presence of the probe materials, quantified as a correction $\delta^{s}$, 
the impurity of the water sample, $\delta^{p}$, the shape of the water sample, $\delta^{b}$, 
and dynamic effects including radiation damping $\delta^{\text{RD}}$ and magnetic 
fields from the precessing proton spins $\delta^{d}$.  In the following we quantify 
each term.   

\subsubsection{Material Effects} \label{sec:mat-perturb}

\begin{figure}[!hbt]
   \centering
   \includegraphics[width=\linewidth]{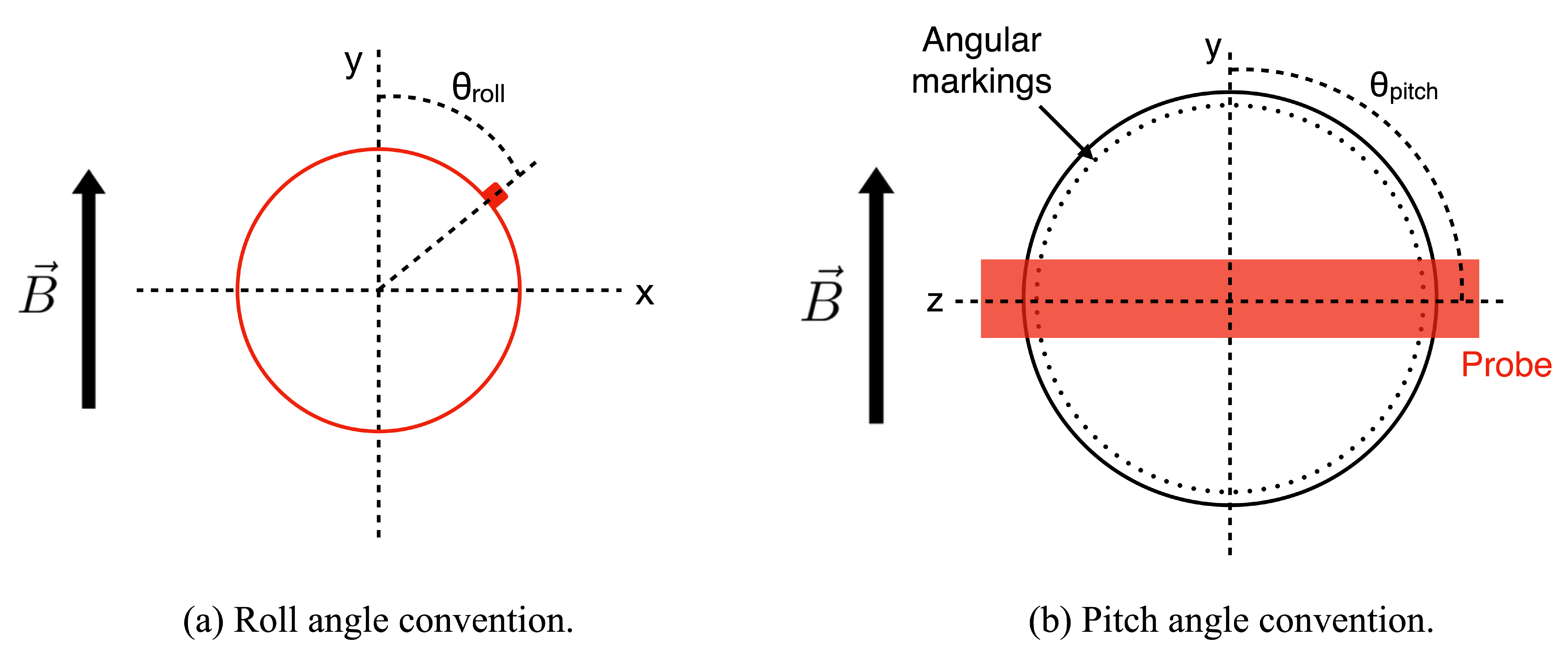}
   \caption{(a) End view of the calibration probe, illustrating the roll angle for material perturbation studies. 
            The roll angle is the angle between the ground screw of the probe (filled square) and the magnetic field axis. 
            (b) Top-down view of the setup used for determining the material perturbation due to the pitch angle of 
            the calibration probe. The probe is mounted on a rotational stage using small pieces of 
            double-sided tape (not shown).  The dotted markings indicate the angular position of the 
            probe axis.  Drawing is not to scale.  
           }
   \label{fig:roll-and-pitch-angles}
\end{figure}

Material effects enter due to the magnetization of all materials surrounding the NMR sample. 
For a perfectly symmetric probe, perturbations arise as the square of the magnetic 
susceptibility and are much smaller than 1\,ppb.  However, the probe is not perfectly 
symmetric; we therefore quantify the material correction $\delta^{s}$ as: 

\begin{equation} \label{eqn:delta-s} 
   \delta^{s}\left( \theta_{\text{roll}},\theta_{\text{pitch}}\right) = \delta^{\text{roll}}\left(\theta_{\text{roll}}\right)  
                                                            + \delta^{\text{pitch}}\left(\theta_{\text{pitch}}\right),
\end{equation} 

\noindent where $\delta^{\text{roll}}$ denotes the correction due to the probe materials 
dependent upon the probe's roll angle $\theta_{\text{roll}}$; that is, how the probe is oriented 
about its long axis relative to the magnetic field axis. 
The term $\delta^{\text{pitch}}$ represents the correction for effects due to the angular 
orientation $\theta_{\text{pitch}}$ of the probe relative to the magnetic field axis.  
These angles are defined in Fig.~\ref{fig:roll-and-pitch-angles}.  

\begin{figure}[!hbt]
   \centering
   \includegraphics[width=\linewidth]{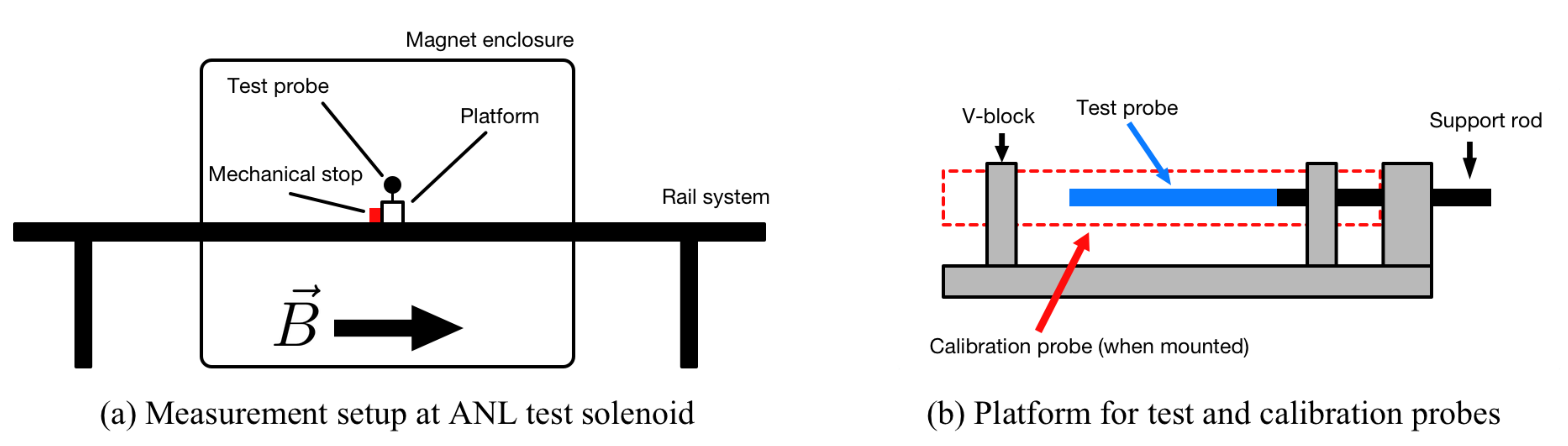}
   \caption{The setup for measuring $\delta^{s}$ for the calibration probe at ANL. 
            (a) The test probe (long axis coming out of the page) is 
            mounted on a retractable platform and secured by a mechanical stop for repeatability of the setup.
            (b) A side view of the platform on which the test probe is mounted.  
            The dotted red rectangle indicates the calibration probe position when mounted on the platform, 
            such that the test probe is inside the calibration probe.  The test probe and calibration 
            probe active volumes are aligned.
           }
   \label{fig:pp-pert-anl}
\end{figure}

The quantity $\delta^{s}$ was determined by comparing measurements inside the calibration probe 
with a fixed test probe and with the calibration probe removed.  The calibration probe was constructed 
specifically so that a test probe can fit inside it.  The test probe was mounted on a rigid stand and locked 
to a single location in the magnet with a repeatability of $<$1\,mm.  This ensures that uncertainties due 
to coupling between 1\,ppb/mm field gradients and alignment errors are limited to be less than 1\,ppb.
Figure~\ref{fig:pp-pert-anl} shows the typical setup.  Three sets of measurements with and without the 
calibration probe were used to correct for linear magnetic field drift in time~\cite{Swanson:2010}.  
The net change with and without the calibration probe is the correction $\delta^{s}$. 


\begin{figure}[!hbt]
   \centering
   \includegraphics[width=\linewidth]{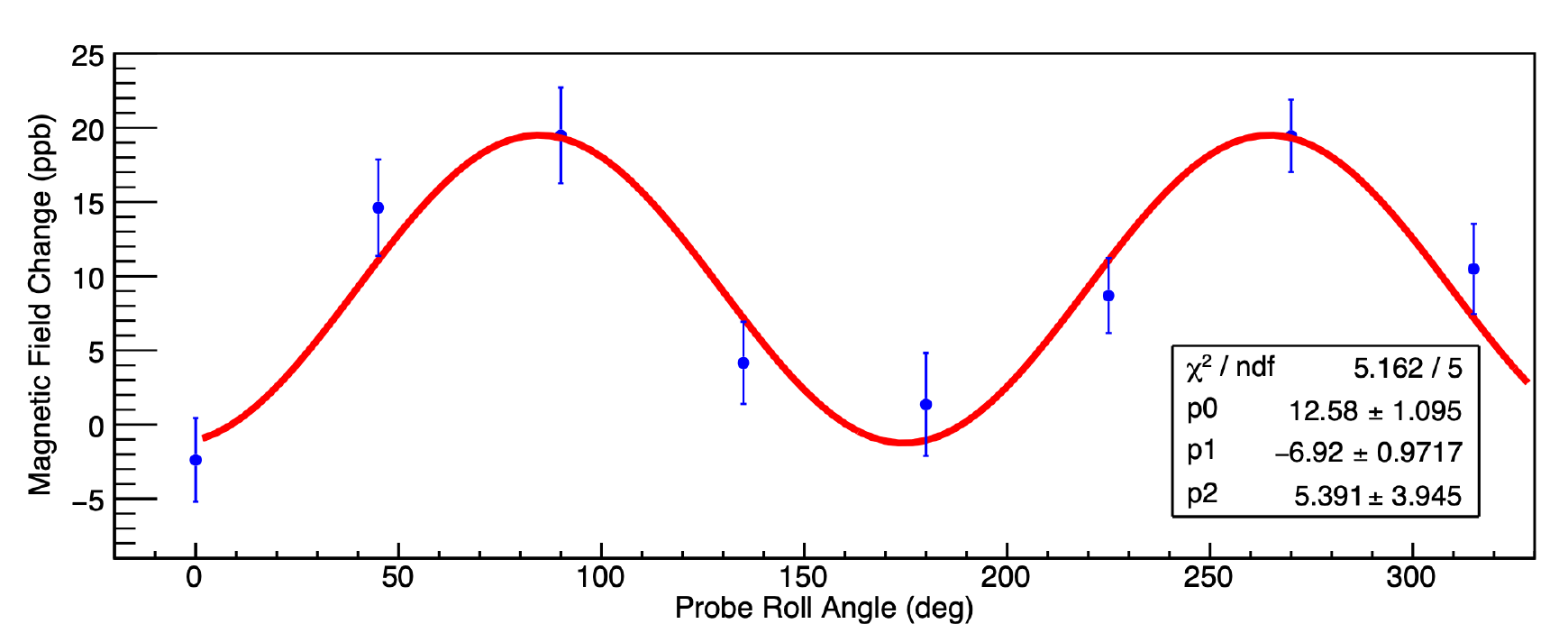}
   \caption{The results of the measured change in the field (in parts per billion) seen by the 
            test probe when the roll angle of the calibration probe is changed.  
            The error bars represent the uncorrelated uncertaintes of the measurements. 
            The red curve represents the fit described in the text. 
            }
   \label{fig:pp-roll-study}
\end{figure}

To quantify $\delta^{s}$ as a function of the probe's roll angle $\theta_{\text{roll}}$, we took 
measurements of the magnetic field of the test solenoid at ANL with the test probe inserted in the calibration probe 
(c.f., Fig.~\ref{fig:pp-pert-anl}).  We compared magnetic field measurements when rotating the calibration 
probe 0$^{\circ}$, $\ldots$, 315$^{\circ}$ in steps of 45$^{\circ}$, see Fig.~\ref{fig:pp-roll-study}.  
The roll angle of 0$^{\circ}$ is defined such that the grounding screw of the probe is parallel to the 
magnetic field.  Overall, we found that the test probe field measurements vary between $-2.4$\,ppb and 19.5\,ppb, 
depending on the roll angle of the calibration probe.  
The data are well described by a simple symmetrical form:

\begin{equation*}  
   f\left(\theta_{r}\right) = p_{0} + p_{1}\left[ 3 \cos^{2}\left( \theta_{r} + p_{2} \right) - 1 \right],  
\end{equation*}

\noindent where $p_{i}$ are determined from the fit to the data.  In particular, we find 
$\delta^{s}(0,0) = (-1.4 \pm 4.0)$\,ppb.  

We must also consider the pitch angle $\theta_{\text{pitch}}$ of the calibration probe relative 
to the external magnetic field axis.  To assess this effect, we mounted the probe on a rotational 
stage that allows for angular rotation of the probe as illustrated in Fig.~\ref{fig:roll-and-pitch-angles}.  
We performed measurements with the probe oriented at angles of 2.5$^{\circ}$ and 5$^{\circ}$ 
relative to its nominal orientation perpendicular to the magnetic field.  The field as measured 
by the calibration probe was lower by ($16.2 \pm 2.5$)\,ppb for the 2.5$^{\circ}$ orientation, and 
lower by ($50 \pm 1$)\,ppb for the 5$^{\circ}$ orientation.  

The results of the studies above are used in Sec.~\ref{sec:fnal-perturb} to quantify the 
perturbations of the calibration probe when installed at Fermilab.  

\subsubsection{Bulk Magnetic Susceptibility} \label{sec:bulk-mag-suscept}  

Possible asymmetries and field perturbations due to the water sample geometry affect the 
bulk magnetic susceptibility $\delta^{b}$ (Eq.~\eqref{eqn:bulk-mag-suscept}).  To evaluate this term 
and its uncertainty, we examine the shape factor.  

The length of the water NMR sample also has an effect on the magnetic field; if the sample were an 
infinitely long cylinder, then no perturbation due to edge effects would be present.  However, 
the water sample has a finite length $L = \left(228.6 \pm 1.0\right)$\,mm, with a diameter 
$D = (4.2065 \pm 0.0065)$\,mm.  To quantify how much the geometry of the sample affects the magnetic 
field, we evaluate the shape factor $\varepsilon$ characterizing the NMR sample magnetization 
as a surface integral~\cite{Hoffman:2006}.  Due to the symmetry of the NMR sample and 
its alignment relative to the magnetic field, the quantity $\varepsilon$ can be expressed as: 

\begin{equation} \label{eqn:eps} 
   \varepsilon \approx \frac{1}{2} \frac{1}{\sqrt{1 + 4R^{2}/L^{2}} }. 
\end{equation}  

\noindent  Evaluating this expression using the values stated above for $L$ and $D = 2R$, we find 
$\varepsilon = 0.499\text{ }915\text{ }37(26)$.  The uncertainty is dominated 
by the uncertainty of the long length $L$ with a sub-dominant contribution 
from the diameter $D$ of the sample tube.  

To confirm the magnetic perturbation due to our water sample length, we took a series of 
measurements retracting the water sample from its seated position in the probe by 5\,mm, 
compared to measurements with the water sample fully inserted in the probe.  We found that 
the field as measured by the calibration probe increases by $(0.2 \pm 0.2)$\,ppb.  This is
consistent with the change in the field expected from evaluating $\varepsilon$ using 
Eq.~\eqref{eqn:eps} for a length $L$ shorter by 5\,mm compared to using the nominal value 
for $\varepsilon$ when inserted into the bulk magnetic susceptibility $\delta^{b}$ (Eq.~\eqref{eqn:bulk-mag-suscept}).  

To determine the uncertainty on $\delta^{b}$, 
we combine the uncertainties on the water magnetic susceptibility $\chi$ (Eq.~\eqref{eqn:water-mag-suscept}) 
and the uncertainty on $\varepsilon$ to obtain an uncertainty of 6.0\,ppb.  The magnitude of 
$\delta^{b}$ will change with temperature; evaluating Eq.~\eqref{eqn:bulk-mag-suscept} at $T = 25^{\circ}$C, 
we determine $\delta^{b} = (-1505.6 \pm 6.0)$\,ppb.  

\subsubsection{Water Sample} \label{sec:water-sample} 

While we utilize an ASTM Type-1 ultra-pure water sample in a highly-symmetric glass tube, we need to quantify 
any magnetic impurities or imperfections that can manifest as perturbations to our magnetic field measurements. 
These are encapsulated in the term $\delta^{p}$, given as: 

\begin{equation*}
   \delta^{p} = \delta^{\text{O}_{2}} + \delta^{w} + \delta^{c}, 
\end{equation*} 

\noindent where $\delta^{\text{O}_{2}}$ denotes the correction due to dissolved oxygen in the water sample;  
$\delta^{w}$ accounts for how much the measured field changes due to a water sample obtained from a different vendor; 
$\delta^{c}$ quantifies the correction due to the camber of the water sample tube (c.f., Sec.~\ref{sec:probe-mech}).  

Oxygen is paramagnetic, and thus can perturb the magnetic field if it is dissolved in the NMR sample.  
We directly measure this effect by preparing a water sample that has been boiled to remove any dissolved 
oxygen, effectively degassing it.  The preparation consisted of heating both the sample tube and the 
water so that when the water was poured into the tube using a syringe, the glass would not shatter.  
We then performed field measurements with the nominal sample with no special preparation and the degassed 
sample, swapping the two samples back and forth.  We measured a total of three trials, where a trial 
consists of a measurement with the nominal sample and the degassed sample. 
We find $\delta^{\text{O}_{2}} = (1.4 \pm 1.0)$\,ppb. 
This result is consistent with expectations. An approximate dissolved oxygen concentration at
25$^{\circ}$C of 8.2\,mg/L corresponds to a number density $N = 1.55 \times 10^{17}$\,cm$^{-3}$ in water.
The volume susceptibility is roughly
$\chi_{\text{O}_{2}} = 4 \pi N (g_{e}\mu_{B})^{2}(S(S+1))/3k_{B}T \approx 1.2 \times 10^{-8}$ where $S = 1$
for oxygen~\cite{Abragam:1961}. This suggests a correction to the shape-dependent bulk susceptibility of
$(1/6) \times \chi_{\text{O}_{2}} \approx 2 \times 10^{-9}$. 

For the term $\delta^{w}$, we studied the change in the NMR frequency when comparing water samples from 
different vendors.  For this test, we prepare two NMR water samples from different vendors and use 
different glass tubes for each sample.  We measure the magnetic field using the calibration probe and 
each of the prepared NMR water samples, and repeat this process for a total of three trials.  
We find $\delta^{w} = (-0.4 \pm 1.4)$\,ppb.  

We did not measure the magnetic perturbation of the water sample glass tube.  This is negligible if the 
tube is infinitely long and perfectly symmetric (i.e., concentric construction with no camber).  
However, the water sample tube has finite length and non-zero camber (c.f., Sec.~\ref{sec:probe-mech}), 
and so we estimate the correction $\delta^{c}$.  We considered how much the measured frequency depends 
on the roll angle of the water sample by taking consecutive measurements using the calibration probe.  
We rotated the water sample by 90$^{\circ}$ for each measurement for a total sample rotation of 360$^{\circ}$.  
Across all measurements, we found that the measured field changed by 1\,ppb, leading to a 
correction $\delta^{c} = (-1 \pm 1)$\,ppb.  

Combining the terms $\delta^{\text{O}_{2}}$, $\delta^{w}$, and $\delta^{c}$ together in quadrature, 
we determined $\delta^{p} = (0 \pm 2)$\,ppb. 

\subsubsection{Radiation Damping} \label{sec:rad-damp} 

Radiation damping arises when the current induced in the RF coil due to the precessing 
proton spins produces its own RF magnetic field that acts to rotate the spins back to being 
pointed along the external magnetic field, artificially reducing the length of the signal. 
This phenomenon is proportional to the relative difference of the resonant frequency 
of the probe $f_{0}$ and the Larmor frequency $f_{L}$ of the proton 
spins.\footnote{For $B = 1.45$\,T, $f_{L} = 61.79$\,MHz.  Our probe is also tuned to this frequency.}  
Radiation damping is also dependent on the $z$ component of the magnetization of the NMR sample 
as a function of time $M_{z}(t)$~\cite{Vlassenbroek:1995}.  To estimate the magnitude of the 
radiation damping effect, we considered the calibration probe's sensitivity to its tune 
and the sensitivity to the change in $M_{z}$.     
 
To assess how much the magnetic field measurement depends on the probe tune, we measured the magnetic 
field at the nominal tune of the ANL test solenoid magnetic field, and compared that to a measurement where we 
detune the probe using its on-board capacitors by $\approx$100\,kHz.  We find no systematic shift larger than 1\,ppb.

We estimated the contribution from the proton spin magnetization $M_{z}(t)$ by varying the 
$\pi/2$ pulse duration and observed how the extracted frequency changed.  We found the effect 
to be less than 2\,ppb over a large range of RF pulse durations from 5--50\,$\mu$s.

For a conservative estimate, we combine the calibration probe's sensitivity to the RF pulse 
duration and the resonant tune in quadrature to find a correction factor $\delta^{\text{RD}} = (0 \pm 3)$\,ppb. 

\subsubsection{Proton Dipolar Field} \label{sec:prot-dipolar} 

Dipolar coupling of spins of distant molecules in the NMR sample can produce 
a non-zero contribution to the measured magnetic field and must be evaluated.  
We estimate the contribution by considering a coupling of a given proton spin 
to a classical dipolar magnetic field due to the other protons as described in 
Ref.~\cite{Jeener:1995jcp}.  We also approximate the effect as a modification 
of the molecular electrons' magnetizations, which changes the bulk magnetic susceptibility 
of oxygen (c.f., Eq.~\eqref{eqn:bulk-mag-suscept}).  
We found the two calculations to be consistent with one another and assigned a 
correction $\delta^{d} = \left( 0 \pm 2.5 \right)$\,ppb. 

\subsection{Effects Specific to the Muon $g-2$ Experiment at Fermilab} \label{sec:fnal-perturb} 

\begin{figure}[!hbt]
   \centering
   \includegraphics[width=\linewidth]{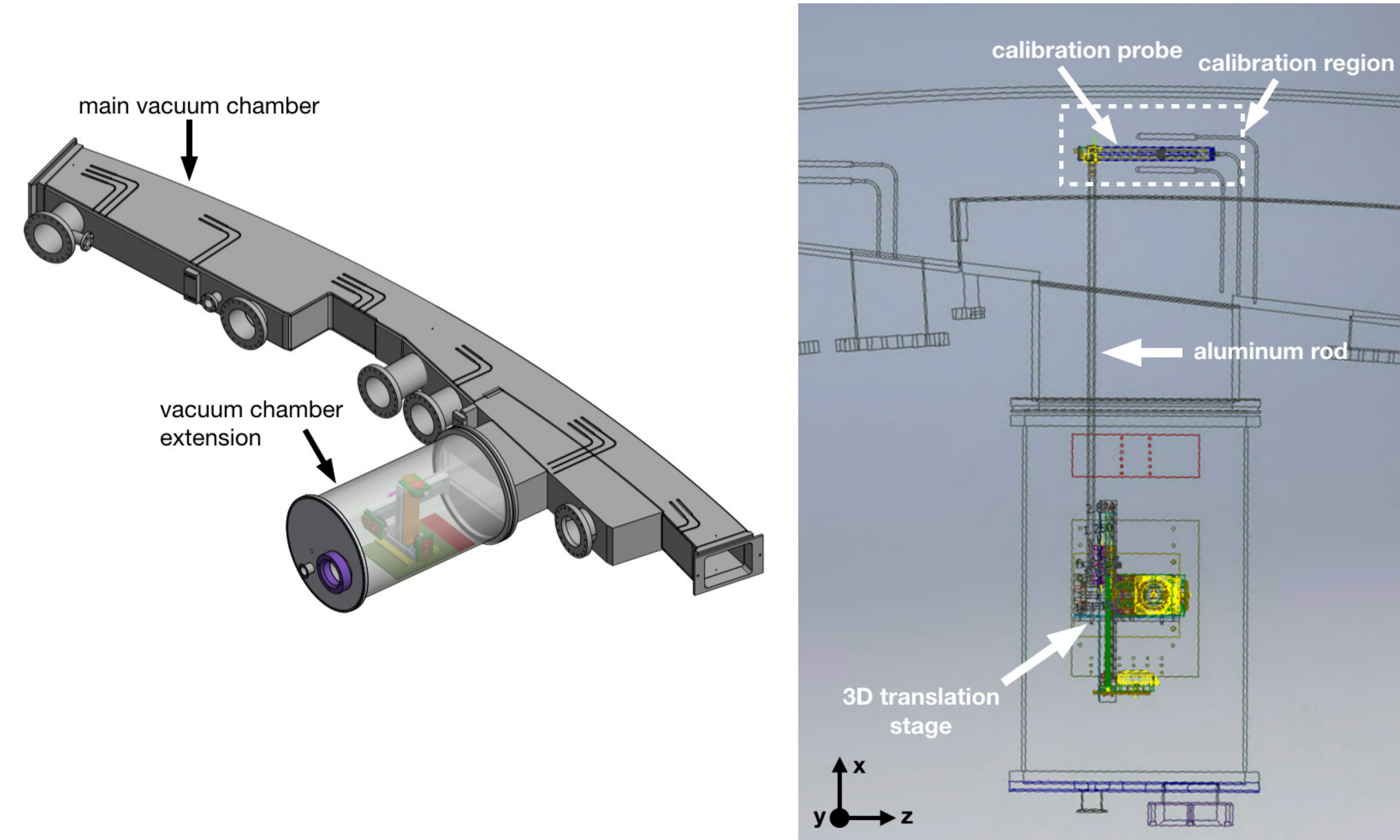}
   \caption{The assembly of the calibration probe on its 3-dimensional translation stage system 
            at Fermilab.  Left: Assembly with respect to the main vacuum chamber of the experiment. 
            Right: Top-down, wireframe view of the assembly. The magnetic field axis is pointed 
            along the $y$ axis.}
   \label{fig:pp-fnal-setup}
\end{figure}

There are perturbations unique to the configuration at Fermilab when the probe was installed 
in-situ for the calibration program with the trolley system. Figure~\ref{fig:pp-fnal-setup} shows the 
configuration at Fermilab. The probe was mounted on a 0.8-m long rod affixed to a 3-dimensional translation 
stage, which allows aligning the calibration probe's sensitive volume to overlap with 
the trolley probes' sensitive volumes.  With this setup, Eq.~\eqref{eqn:delta-s} has to be 
modified to:

\begin{equation} \label{eqn:delta-sf} 
   \delta^{s,\text{config}} = \delta^{s}(0,0) + \delta^{\text{mag}} + \delta^{\text{roll}} + \delta^{\text{pitch}} 
                            + \delta^{\text{cable}} + \delta^{\text{vac}} + \delta^{T},
\end{equation} 

\noindent where we use $\delta^{s}(0,0)$ (Eq.~\eqref{eqn:delta-s}) since the probe is 
installed at Fermilab with $\theta_{\text{roll}} = \theta_{\text{pitch}} \approx 0^{\circ}$; 
$\delta^{\text{mag}}$ is the correction due to magnetic images of the probe induced in the 
$g-2$ magnet pole pieces; $\delta^{\text{roll}}$ denotes the correction due to a non-zero roll 
angle of the probe; $\delta^{\text{pitch}}$ is due to a non-zero pitch angle 
of the probe; $\delta^{\text{cable}}$ denotes the correction due to the SMA cable that delivers 
the $\pi/2$ signal and receives the FID signal;  $\delta^{\text{vac}}$ denotes the correction 
factor to account for measurements being performed in air as opposed to in vacuum.  
In the following, we address each term.  Note that $\delta^{s}(0,0)$ and $\delta^{\text{mag}}$ have 
been evaluated together as a single term, see Sec.~\ref{sec:mag-img}.
 
\subsubsection{Magnetic Images} \label{sec:mag-img} 

\begin{figure}[!hbt]
   \centering
   \includegraphics[scale=0.25]{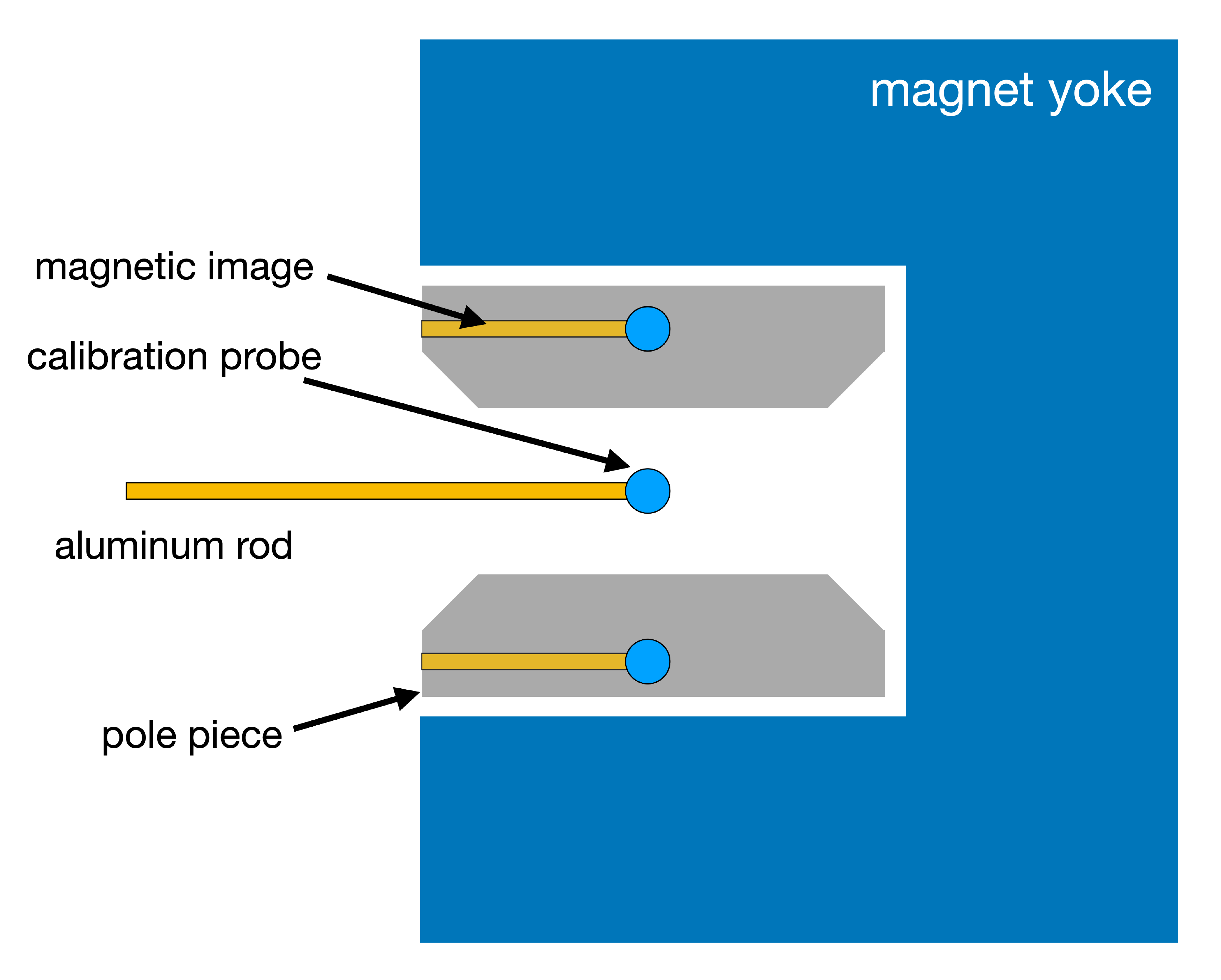}
   \caption{An illustration of magnetic images of the calibration probe and its aluminum rod 
            in the steel pole pieces of the Muon $g-2$ storage ring magnet. }
   \label{fig:mag-img-cartoon}
\end{figure}

When the probe is placed in the magnetic storage ring at Fermilab, it induces magnetic images 
in the upper and lower pole pieces, see Fig.~\ref{fig:mag-img-cartoon}.  To determine the correction 
due to the calibration probe materials and magnetic images, $\delta^{s}(0,0) + \delta^{\text{mag}}$, 
we need to evaluate the correction for the probe oriented with the roll and pitch angle of $0^{\circ}$ 
inside the muon storage ring magnet.  We accomplish this by measuring the image effect at ANL in the test solenoid.  
To check our results, we also measured the effect in-situ at Fermilab. 

To measure the magnetic image effect in the test solenoid at ANL, we measured the change in the 
magnetic field seen by a test probe with and without the presence of the calibration probe placed one 
image distance (18\,cm) away.  We performed six trials of test probe readings with and 
without the calibration probe placed at its image location.  We find the calibration probe increases 
the test probe magnetic field reading by $(3.0 \pm 2.7)$\,ppb.  In the Muon $g-2$ magnet, there is a 
magnetic image above and below the calibration probe location; by symmetry, this doubles the magnitude 
of the measured effect at ANL, resulting in a magnetic image of $(6.0 \pm 5.4)$\,ppb.  
The measured value of the probe material perturbation for $\theta_{\text{roll}} = \theta_{\text{pitch}} = 0^{\circ}$ 
is $\delta^{s}(0,0) = (-1.4 \pm 4.0)$\,ppb (Sec.~\ref{sec:mat-perturb}).  The perturbation of the adapter was 
measured separately to be $(-8.0 \pm 3.5)$\,ppb. Combining these terms together, we determine the 
correction $\delta^{s}(0,0) + \delta^{\text{mag}} = \left( 3.4 \pm 7.6 \right)$\,ppb.  

\begin{figure}[!hbt]
   \centering
   \includegraphics[scale=0.20]{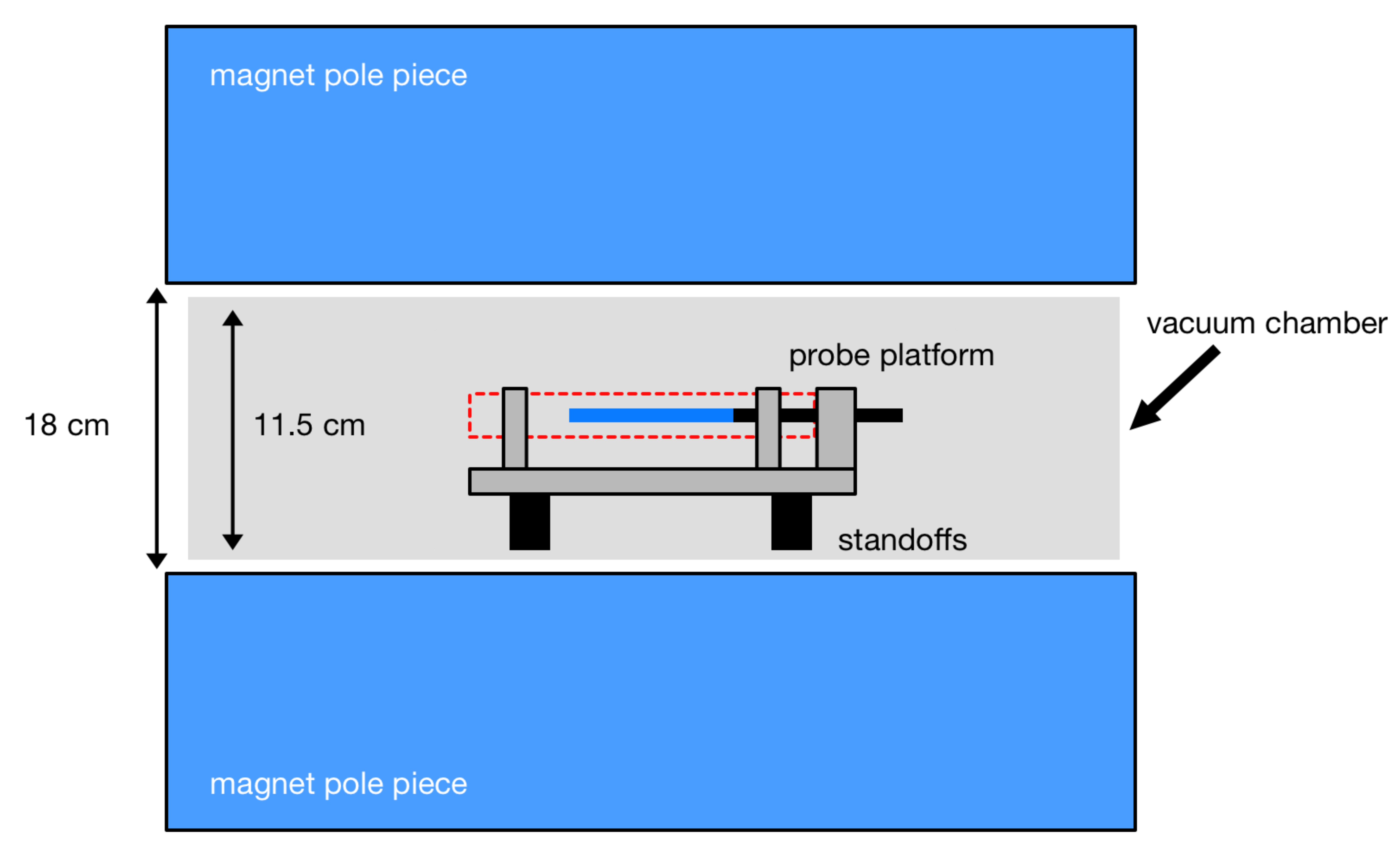}
   \caption{The setup at Fermilab for the measurement of the magnetic images of the calibration probe 
            induced in the magnet pole pieces.  The dotted red rectangle indicates the calibration probe 
            position when mounted.  The probe's adapter that connects it to its support rod was included 
            in this study (not drawn).  The pole-pole separation of 18\,cm is indicated.  
            The vacuum chamber that fits in between the pole pieces is represented by the large 
            light gray band between the magnet pole pieces.  The clearance into this enclosure is 
            roughly 11.5\,cm.  Drawing is not to scale.}
   \label{fig:pp-images}
\end{figure}

The general setup at Fermilab is shown in Fig.~\ref{fig:pp-images}.  The measurements were 
performed at the center of the magnet gap using the same procedure discussed in Sec.~\ref{sec:mat-perturb}. 
This procedure was repeated for 8 pairs of measurements 
with the calibration probe installed and not installed on the stage.  To correct for field drift, 
we stationed the trolley roughly 124\,cm downstream in azimuth in the storage ring to take data 
simultaneously with our test probe measurements.  Averaging over all trials, we find good agreement with 
the value measured at ANL.  The magnetically noisier environment at Fermilab makes it very difficult to obtain 
results with uncertainties smaller than 8\,ppb. 

To build confidence in our measurements, we also calculate the magnetic images of the probe and 
its support rod numerically. The magnetic image of an object in a nearby material 
with relative magnetic permeability $\mu_{r}$ is written~\cite{Jackson:1999}: 

\begin{equation*}
   \Delta B' \approx \left( \frac{\mu_{r}-1 }{\mu_{r}+1} \right) \Delta B \left(x,y,z'\right) 
             \approx \Delta B \left(x,y,z' \right). 
\end{equation*}

\noindent This approximation follows because $\mu_{r} = 1450$ for the ultra-pure magnet steel at 
1.45\,T~\cite{Danby:2001eh}.  As such, to evaluate the magnetic image of the material, we compute 
the magnetic perturbation of the calibration probe at a vertical height $z'$ that describes the 
distance from the probe active volume to the image location in a given pole piece.  
This calculation is performed for both the upper and lower pole piece, where the calibration probe is 
located at the center of the magnet gap.  
To determine the magnetic image of the probe, 
we compute the magnetic image of each component of the probe, following: 

\begin{equation} \label{eqn:pert-calc} 
   \Delta \vec{B} \left( \vec{r} \right) = \int_{V} d^{3}\vec{r}\,' \left[ 
\frac{ 3 \left[ \vec{m}\left( \vec{r}\,'\right) \cdot \left(\vec{r}-\vec{r}\,'\right) \right] \left( \vec{r} - \vec{r}\,'\right) }{\vert\vec{r}-\vec{r}\,'\vert^{5}} 
- \frac{\vec{m}\left( \vec{r}\,' \right)}{\vert \vec{r}-\vec{r}\,' \vert^{3}} \right],
\end{equation} 

\noindent where the magnetic moment $d\vec{m} = \chi \vec{B}_{\text{ext}}\left(\vec{r}\,'\right) dV$ 
for a given material.  We sum over all materials in the probe, including its adapter that connects the 
probe to the support rod; we do not include the effect due to the water and 
its glass tube, which we estimate to be small.  
The computed effect is consistent with our measurements at ANL. 

We evaluate the magnetic perturbation and image of the calibration probe support rod separately 
because of its very different geometry compared to the calibration probe. 
We did not make a measurement at ANL since the rod is too large to fit into the test magnet; due to the 
magnetically-noisy environment at Fermilab, measurements were difficult to perform.  
Following the same general prescription above, we compute
$\left( \delta^{s}(0,0) + \delta^{\text{mag}} \right)_{\text{calc}}^{\text{rod}} = \left( 8.0 \pm 5.0 \right)$\,ppb.  
We used the commercial software OPERA\footnote{\url{http://operafea.com}.} for finite-element calculations of the 
magnetic field to account for the field falloff near the pole piece edges.  

To evaluate the term $\delta^{s}(0,0) + \delta^{\text{mag}}$ for the calibration probe and support rod, we combine 
the measurement at ANL for the probe and the calculation for the support rod and find 
$\delta^{s}(0,0) + \delta^{\text{mag}} = \left( 11.4 \pm 9.0 \right)$\,ppb.

\subsubsection{Probe Roll Angle}

As discussed in Sec.~\ref{sec:mat-perturb}, the probe roll angle may change the size of the correction; 
as such, we must use a specific value from our measurements based on what the roll angle is in the installed 
setup at Fermilab.  The probe is installed such that its ground screw (see Fig.~\ref{fig:pp-assembly}) is 
pointed along the magnetic field, and we have measured the roll angle of the probe to be $\ll$1$^{\circ}$ under 
the roll angle convention adopted for the study.  Therefore, we give a conservative estimate of the effect 
due to this asymmetry to be $\delta^{\text{roll}} = (0 \pm 1)$\,ppb based on our results in Sec.~\ref{sec:mat-perturb}. 

\subsubsection{Probe Pitch Angle} 

The pitch of the calibration probe when installed in-situ at Fermilab was measured to be 0.68$^{\circ}$.  
We therefore make a conservative estimate that the pitch angle acts to reduce the field by 
$\delta^{\text{pitch}} = (4.4 \pm 4.4)$\,ppb, where we linearly scale down the value measured at 
the 2.5$^{\circ}$ angle at ANL (Sec.~\ref{sec:mat-perturb}).    

\subsubsection{SMA Cable} 

The magnetic properties of the SMA cable that connects to the probe must also be considered.  
We measured its perturbation in the ANL test solenoid by placing the SMA connector roughly 12.7\,cm 
from the active volume of our test probe.  We maintain the same distance and relative orientation 
between the SMA connector and the active volume of the calibraton probe.
Comparing field measurements with the cable set up and with it removed, we find that the field 
is lowered due to the presence of the cable.  The corresponding correction value is 
$\delta^{\text{cable}} = (1.4 \pm 3.0)$\,ppb.  

\subsubsection{Vacuum Effect} \label{sec:vac-effect} 

When the studies in Secs.~\ref{sec:mat-perturb} and~\ref{sec:mag-img} were conducted, 
the results effectively describe the field perturbation of the calibration probe relative to an 
equivalent calibration probe made of air.  In the Muon $g-2$ Experiment, the perturbation of the 
calibration probe must be known in a vacuum environment.  

To evaluate this effect, we compute the perturbation (Eq.~\ref{eqn:pert-calc}) for a hypothetical 
calibration probe that has a magnetic susceptibility defined as 
$\chi' = \chi - 0.2\chi_{\text{O}_{2}}$,\footnote{The contribution due to the magnetic 
susceptibility of nitrogen is considerably smaller.} where $\chi_{\text{O}_{2}}$ is the 
magnetic susceptibility of oxygen.  We find that the perturbation of the probe turns out to 
be 2\,ppb smaller when evaluating Eq.~\eqref{eqn:pert-calc} using $\chi'$ for each material.  
In the Muon $g-2$ analysis, we account for this using a correction 
$\delta^{\text{vac}} = \left( -2 \pm 2 \right)$\,ppb. 
This correction is accounted for in our calculation of the magnetic perturbation and image effect in 
Sec.~\ref{sec:mag-img}.

\subsubsection{Temperature Effect}

When extracting the shielded proton precession frequency from NMR measurements, we account for the 
temperature dependence of the diamagnetic shielding of water (c.f., Eq.~\eqref{eqn:water-diamag-shield}). 
As such, the stability of our temperature readout is considered.  The stability of 0.5$^{\circ}$C 
as described in Sec.~\ref{sec:design} results in a temperature correction $\delta^{T} = (0 \pm 5)$\,ppb.

\subsubsection{Summary of Probe Perturbations and Uncertainties at Fermilab} \label{sec:pp-pert-summary} 

The results of the studies above are given in Table~\ref{tab:pp-pert-fnal-results}.  The material correction 
of the probe in-situ at Fermilab has been measured to be: 

\begin{equation*}
   \delta^{s,\text{config}} = \delta^{s}(0,0) + \delta^{\text{mag}} + \delta^{\text{roll}} + \delta^{\text{pitch}} 
                            + \delta^{\text{cable}} + \delta^{\text{vac}} + \delta^{T} = (15.2 \pm 12.0) \text{ ppb}.  
\end{equation*} 

\begin{table}[!hbt]
\centering
\caption{The calibration probe magnetic corrections due to the probe materials and other effects 
         specific to the setup at Fermilab.  The total uncertainty in the final row is the quadrature 
         sum of each uncertainty.
         }
\label{tab:pp-pert-fnal-results}
\small{\begin{tabular}{|c|c|c|c|}
\hline
{\bf Source} & {\bf Symbol} & {\bf Magnitude (ppb) } & {\bf Uncertainty (ppb)}\\
\hline
Material + Images                    & $\delta^{s}(0,0) + \delta^{\text{mag}}$                         & $11.4$  & $9.0$   \\ \hline
Roll Angle                           & $\delta^{\text{roll}} (\theta_{\text{roll}} = 0^{\circ})$       & $0$     & $1$     \\ \hline
Pitch Angle                          & $\delta^{\text{pitch}} (\theta_{\text{pitch}} = 0.68^{\circ})$  & $4.4$   & $4.4$   \\ \hline
SMA Cable                            & $\delta^{\text{cable}}$                                         & $1.4$   & $3.0$   \\ \hline
Vacuum Effect                        & $\delta^{\text{vac}}$                                           & $-2$    & $2$     \\ \hline
Water Sample Temperature             & $\delta^{T}$                                                    & $0$     & $5$     \\ \hline
{\bf Total Material Correction}      & $\delta^{s,\text{config}}$                                      & $15.2$  & $12.0$  \\ \hline
\end{tabular}}
\end{table}

\noindent This includes effects due to the probe's SMA cable and its support rod.  We additionally 
account for the presence of oxygen in our measurements. 

For the Muon $g-2$ Experiment, the probe's magnetic corrections sum to (at $T = 25^{\circ}$C):

\begin{equation*}
   \delta^{t,\text{config}} = \delta^{s,\text{config}} + \delta^{p} + \delta^{\text{RD}} + \delta^{d}  
                            = (15.2 \pm 12.7) \text{ ppb}.
\end{equation*}

\noindent The correction due to water sample paramagnetic impurities is $\delta^{p} = (0 \pm 2)$\,ppb 
(Sec.~\ref{sec:water-sample}).  The radiation damping term is estimated to be 
$\delta^{\text{RD}} = (0 \pm 3)$\,ppb (Sec.~\ref{sec:rad-damp}), while the contributions to the dipole 
field due to the precessing protons in the water are estimated as $\delta^{d} = (0 \pm 2.5)$\,ppb 
(Sec.~\ref{sec:prot-dipolar}).  These numbers are summarized in Table~\ref{tab:pp-pert-results}.  
Combining the uncertainty on the probe total correction $\delta^{t,\text{config}}$ with that of the 
bulk magnetic susceptibility $\delta^{b}$, the calibration probe is accurate to 15\,ppb in extracting 
the shielded proton frequency $\omega_{p}'$ for the Muon $g-2$ Experiment at Fermilab.  

\begin{table}[!hbt]
\centering
\caption{The calibration probe magnetic corrections due to intrinsic properties $\delta^{p}$, $\delta^{\text{RD}}$, 
         and $\delta^{d}$, and configuration-specific effects $\delta^{s,\text{config}}$ relevant to the Muon $g-2$ 
         Experiment at Fermilab.  The contribution from the bulk magnetic susceptibility $\delta^{b}$ is also 
         listed and evaluated at 25$^{\circ}$C.  The last row shows the total correction, $\delta^{b} + \delta^{t,\text{config}}$, 
         evaluated at 25$^{\circ}$C.  
         }
\label{tab:pp-pert-results}
\small{\begin{tabular}{|c|c|c|c|}
\hline
{\bf Source} & {\bf Symbol} & {\bf Magnitude (ppb) } & {\bf Uncertainty (ppb)}\\
\hline
Material Correction (Fermilab)   & $\delta^{s,\text{config}}$                                          & $15.2$    & $12.0$  \\ \hline
Water Impurity                   & $\delta^{p}$                                                        & $0$       & $2$     \\ \hline
Radiation Damping                & $\delta^{\text{RD}}$                                                & 0         & $3$     \\ \hline  
Proton Dipolar Field             & $\delta^{d}$                                                        & 0         & $2.5$   \\ \hline
Bulk Magnetic Susceptibility     & $\delta^{b} \left(\text{H}_{2}\text{O},T=25^{\circ}\text{C}\right)$ & $-1505.6$ & $6.0$   \\ \hline 
{\bf Total Correction}           & --- & $-1490.4$ & $14.1$   \\ \hline 
\end{tabular}}
\end{table}

\section{Cross Checks} \label{sec:cross-check} 

In addition to quantifying the various correction terms $\delta$ for the 
calibration probe, we have performed extensive cross-checks against other 
calibration standards used for muon $g-2$ experiments; in particular, the E821 spherical 
water probe~\cite{Fei:1997sd}, the newly-constructed $^{3}$He 
probe~\cite{Farooq:2019phd,Farooq:2020swf}, and a water-based continuous-wave (CW) 
NMR probe to be used for the future J-PARC muon $g-2$ experiment~\cite{Yamaguchi:2019iwm}.  
All cross-check measurements were performed in the test magnet at ANL.  

\subsection{Comparison to E821}

We performed a direct comparison against the E821 probe, using a platform 
similar to the one shown in Fig.~\ref{fig:pp-pert-anl}.  The calibration 
probe was placed on the stage and measured the magnetic field at the center 
of the solenoid; the E821 probe was swapped into the same location and the 
measurement for that probe was also performed.  We conducted three pairs of measurements.
The measured difference between the spherical (E821) and cylindrical (calibration) probes 
was $\delta^{b}_{\text{sph.}-\text{cyl.}} = (1514 \pm 15)$\,ppb, 
where the uncertainty is dominated by the asphericity of the E821 water sample.     
This difference is in agreement with Eq.~\eqref{eqn:bulk-mag-suscept}, which 
gives 1506\,ppb for the water magnetic susecptibility 
$\chi\left( \text{H}_{2}\text{O},T = 25^{\circ}\text{C}\right) = -9038 \times 10^{-9}$ 
and the shape factor $\varepsilon = 1/3$ for a perfect sphere and 1/2 for 
an infinite cylinder with its long axis perpendicular to $\vec{B}$. 

\subsection{Comparison to $^{3}$He}

The cross check against the $^{3}$He probe is sensitive to very different systematic 
effects due to the very different probe constructions and different NMR samples. 
A similar measurement scheme to the one presented in the previous section was used to 
compare the $^{3}$He and E821 probes. This work is described in Ref.~\cite{Farooq:2019phd}. 
After applying corrections for the material perturbations of the E821 probe and $^{3}$He probe, 
and correcting the E821 probe to $T = 25^{\circ}$C, the ratio of the $^{3}$He to  
the E821 probe frequencies was measured to be $0.761\text{ }786\text{ }139(29)$ (38\,ppb). 
This agrees with a previous measurement of the ratio of frequencies from $^{3}$He 
and water in a spherical sample, $0.761\text{ }786\text{ }1313(33)$ (4.3\,ppb)~\cite{Flowers:1993}. 
This result indirectly calibrates the calibration probe to the $^{3}$He probe 
via the E821 probe; the calibration probe is validated to $(10 \pm 38)$\,ppb. 

\subsection{Comparison to the J-PARC Calibration Probe}

The magnetic field team at J-PARC has built a water-based CW-NMR 
probe~\cite{Yamaguchi:2019iwm} for the future muon $g-2$ experiment at J-PARC~\cite{Torii:2015sra}.  
We have undertaken an extensive cross-calibration program comparing our pulsed 
NMR probe with their CW NMR probe.  While both probes have similar construction materials 
and geometries, the NMR measurement approach is very different, and thus sensitive to 
different systematic effects.  The cross-calibration program was carried out at 
$\vert \vec{B} \vert = 1.45$\,T and 1.7\,T, where we have constructed an additional calibration 
probe to function at 1.7\,T.  The analysis of the data from that 
program is ongoing.  A future cross-calibration program is planned at 3\,T, for which 
we have constructed a calibration probe.

\section{Conclusion} \label{sec:conclusion} 

We have presented the design, performance, and magnetic characteristics and their 
associated uncertainties for a highly accurate water-based NMR 
calibration magnetometer. The probe has demonstrated a single-FID resolution of 
better than 100\,ppt in a highly-uniform test solenoid at ANL; in-situ at Fermilab,  
the probe has a resolution of 10\,ppb.   
 
The probe's intrinsic magnetic characteristics $\delta^{s}$, $\delta^{p}$, $\delta^{\text{RD}}$, 
and $\delta^{d}$ were studied and quantified.  In the probe's deployment in the 
Muon $g-2$ Experiment at Fermilab to calibrate the trolley probe magnetic field 
measurements, configuration-specific perturbations $\delta^{s,\text{config}}$ were 
quantified.  The probe was found to be accurate to 15\,ppb in extracting the 
shielded-proton Larmor precession frequency $\omega_{p}'$ from the NMR magnetic 
field measurements, exceeding the 35\,ppb goal for the experiment. 

We have performed a careful validation program comparing the calibration probe 
to the E821 probe and to a novel $^{3}$He-based probe that has been 
recently developed.  We found excellent agreement with the BNL probe after 
accounting for Eq.~\eqref{eqn:bulk-mag-suscept}, and via an indirect approach, 
found agreement with the $^{3}$He probe at the $(10 \pm 38)$\,ppb level.

The absolute magnitude of the magnetic field may be extracted using the calibration 
probe when accounting for the water diamagnetic shielding term $\sigma$.  
Incorporating the uncertainty on $\sigma$ of 11\,ppb~\cite{Mohr:2012tt}, the calibration probe is accurate 
with a precision of 18.6\,ppb in determining the free-proton Larmor precession frequency 
from water NMR measurements.   
 
\section{Acknowledgements} 

We would like to thank our colleague K.~Sasaki for many useful discussions, 
and R.~Reimann for carefully reading the manuscript and providing helpful comments.  
We gratefully acknowledge the engineering and technical support received at both 
Argonne National Lab and Fermilab in the process of building the probe and executing 
the various measurement programs.  This work is supported by the U.S. Department of 
Energy, Office of High Energy Physics under contracts DE-FG02-88ER40415 (University of Massachusetts),
DE-AC02-06CH11357 (Argonne National Laboratory), and DE-FG02-97ER41020 (University of Washington), 
Department of Energy, Office of Nuclear Physics under contract DE-AC05-06OR23177 
(Thomas Jefferson National Accelerator Facility), and NSF grant PHY-1812314 (University of Michigan).

   
\providecommand{\href}[2]{#2}\begingroup\raggedright\endgroup

   \appendix 
\section{Component Listings} \label{sec:comp-listings}  

In this section we provide tables of the various components and their corresponding 
manufacturer and part number details.  Table~\ref{tab:probe-parts} contains the 
parts for the calibration probe and Tables~\ref{tab:spu-parts} and~\ref{tab:daq-parts} 
contains the parts used in the data acquisition system. 

\begin{table}[!hbt]
\centering
\caption{Components in the NMR calibration probe.
         }
\label{tab:probe-parts}
\footnotesize{\begin{tabular}{|p{0.12\linewidth}|p{0.18\linewidth}|p{.15\linewidth}|p{.55\linewidth}|}
\hline
{\bf Part} & {\bf Manufacturer } & {\bf Model } & {\bf Link} \\
\hline
RF coil               & Doty Scientific\texttrademark{} & 90179 
                      & \url{https://dotynmr.com} \\ \hline
15-mm OD glass tube   & Wilmad LabGlass\texttrademark{} & 515-7PP-9 
                      & \url{https://customglassparts.com} \\ \hline
5-mm OD glass tube    & Wilmad LabGlass\texttrademark{} & 507-PP-9 
                      & \url{https://customglassparts.com} \\ \hline
Water sample          & Cole Parmer\texttrademark{} & 8005496 
                      & \url{https://coleparmer.com/i/labchem-water-deionized-acs-grade-astm-type-i-1-l/8005496} \\ \hline
Capacitors            & Knowles Voltronics\texttrademark{} & NMA1J12HVS
                      & \url{https://www.digikey.com/en/products/detail/knowles-voltronics/NMA1J12HVS/6362741} \\ \hline
Temperature sensor    & TE Connectivity\texttrademark{} & NB-PTCO-165 
                      & \url{https://www.te.com/usa-en/product-NB-PTCO-165.html} \\ \hline
\end{tabular}}
\end{table}

\begin{table}[!hbt]
\centering
\caption{Components in the NMR data acquistion system.  The column Diagram Label refers to those
         in Fig.~\ref{fig:spu-schematic}.
         }
\label{tab:spu-parts}
\footnotesize{\begin{tabular}{|p{0.12\linewidth}|p{0.18\linewidth}|p{.15\linewidth}|p{.55\linewidth}|}
\hline
{\bf Diagram Label} & {\bf Manufacturer } & {\bf Model } & {\bf Link} \\
\hline
F1 and F2   & Stanford Research Systems\texttrademark{} & SG380
                      & \url{https://www.thinksrs.com/products/sg380.html} \\ \hline
A1          & Tomco\texttrademark{}           & BT00250-Gamma
                      & \url{https://www.everythingrf.com/products/microwave-rf-amplifiers/tomco-technologies/567-503-bt00250-gamma} \\ \hline

SW          & Mini-Circuits\texttrademark{}    & ZSW2-63DR+
                      & \url{https://www.minicircuits.com/WebStore/dashboard.html?model=ZSW2-63DR\%2B} \\ \hline
A2          & Pasternack\texttrademark{}       & 15A1013
                      &  \url{https://www.pasternack.com/50-db-gain-1000-mhz-low-noise-high-gain-amplifier-sma-pe15a1013-p.aspx} \\ \hline
BP          & Lark Engineering\texttrademark{} & Custom
                      & \url{https://www.bench.com/lark} \\ \hline
MX          & Mini-Circuits\texttrademark{}    & ZAD-3H+
                      & \url{https://www.minicircuits.com/WebStore/dashboard.html?model=ZAD-3H\%2B} \\ \hline
LP1 and LP2 & KR Electronics\texttrademark{} & Custom
                      & \url{http://www.krelectronics.com} \\ \hline
OA          & Analog Devices\texttrademark{}   & AD797
                      & \url{https://www.analog.com/en/products/ad797.html} \\ \hline
DG          & Struck Innovative Systeme\texttrademark{} & SIS3316
                      & \url{https://www.struck.de/sis3316.html} \\ \hline
\end{tabular}}
\end{table}

\begin{table}[!hbt]
\centering
\caption{Additional components in the NMR data acquisition system.  Also included are the power supplies 
         used in the power supply unit.  These items are not shown in Fig.~\ref{fig:spu-schematic}.  
         }
\label{tab:daq-parts}
\footnotesize{\begin{tabular}{|p{0.12\linewidth}|p{0.18\linewidth}|p{.15\linewidth}|p{.55\linewidth}|}
\hline
{\bf Part} & {\bf Manufacturer } & {\bf Model } & {\bf Link} \\
\hline
FPGA                    & Acromag\texttrademark{} & IP-EP201 
                        & \url{https://www.acromag.com/shop/embedded-i-o-processing-solutions/pcie-products/pcie-carrier-boards/mezzanine-i-o-modules-for-pcie-carriers/fpga-i-o-support-for-pcie-carriers/ip-ep200-cyclone-ii-fpga-with-digital-i-o-jtag-configured/?attribute\_part-number=IP-EP201\%3A+48+TTL+bidirectional+I\%2FO} \\ \hline
VME crate               & Wiener\texttrademark{} & 6U VME 6023 
                        & \url{https://www.wiener-d.com/product/6u-vme64x-6023-full-size-chassis/} \\ \hline
Digital multimeter      & Keithley\texttrademark{} & 2100 
                        & \url{https://www.tek.com/en/products/keithley/digital-multimeter/2100-series} \\ \hline
3.3\,V power supply     & Acopian\texttrademark{} & 
                        & \url{https://www.acopian.com/single-l-screw-m.html} \\ \hline 
$\pm$5\,V power supply  & Bel Power Solutions\texttrademark{} & HAA5-1.5/OVP-AG 
                        & \url{https://belfuse.com/product/part-details?partn=HAA5-1.5/OVP-AG} \\ \hline 
12\,V power supply      & Bel Power Solutions\texttrademark{} & HB12-1.7-AG 
                        & \url{https://belfuse.com/product/part-details?partn=HB12-1.7-AG} \\ \hline 
\end{tabular}}
\end{table}

\end{document}